\documentclass[]{pasj02} 
\usepackage[switch,mathlines]{lineno} 

\jyear{2025}
\Received{2025/10/21}
\Accepted{2026/1/27}

\graphicspath{{./}{figures/}} 

\usepackage{comment}
\usepackage{ulem}
\usepackage{url}

\begin{document} 

\title{Self-consistent \textit{N}-body simulation of Planetesimal-Driven Migration. II. The effect of PDM on planet formation from a planetesimal disk}

\author{
 Tenri \textsc{Jinno},\altaffilmark{1}\altemailmark\orcid{0009-0001-5384-4202} \email{223s415s@gsuite.kobe-u.ac.jp} 
 Takayuki R. \textsc{Saitoh},\altaffilmark{1,2}\altemailmark\orcid{0000-0001-8226-4592}
 \email{saitoh@people.kobe-u.ac.jp} 
 Yoko \textsc{Funato},\altaffilmark{3}\altemailmark\orcid{0000-0002-6992-7600} \email{funato@system.c.u-tokyo.ac.jp}
 and 
 Junichiro \textsc{Makino}\altaffilmark{1,2}\altemailmark\orcid{0000-0002-0411-4297} \email{jmakino@people.kobe-u.ac.jp}
}
\altaffiltext{1}{Department of Planetology, Graduate School of Science, Kobe University, 1-1 Rokkodai-cho, Nada-ku, Kobe, Hyogo 657-8501, Japan}

\altaffiltext{2}{Center for Planetary Science (CPS), Graduate School of Science, Kobe University, 1-1 Rokkodai-cho, Nada-ku, Kobe, Hyogo 657-8501, Japan}

\altaffiltext{3}{General Systems Studies, Graduate School of Arts and Sciences, The University of Tokyo, Komaba, Meguro, Tokyo, 153-8902, Japan}




\maketitle

\begin{abstract}
According to the canonical planet formation theory, planets form ``in-situ'' within a planetesimal disk via runaway and oligarchic growth. This theory, however, cannot naturally account for the formation timescale of ice giants or the existence of diverse exoplanetary systems. Planetary migration is a key to resolving these problems. One well-known mechanism of planetary migration is planetesimal-driven migration (PDM), which can let planets undergo significant migration through gravitational scattering of planetesimals. In our previous paper (\cite{2024PASJ...76.1309J}, PASJ, 76, 1309), we investigated the migration of a single planet through PDM, addressing previously
unexplored aspects of both the gravitational interactions among planetesimals and the interactions with disk gas. Here we perform the first high-resolution simulations of planet formation from a large-scale planetesimal disk, incorporating planet-gas disk interactions, planet-planetesimal interactions, gravitational interactions among all planetesimals, and physical collisions between planetesimals to investigate the role of PDM in the planet formation process. Our results show that protoplanets undergo dynamic inward/outward migrations during the runaway growth stage via PDM. Moreover, orbital repulsion combined with PDM tends to make two groups of protoplanets, outer ones going outward and inner ones going inward. Such dynamic migration significantly influences the early stages of planetary formation. These findings provide a viable pathway for the formation of Earth-like planets and ice giants' cores. Furthermore, they suggest that a standard protoplanetary disk model can account for the planetary migration necessary to explain diverse exoplanetary systems without the need for additional hypotheses.
\end{abstract}


\section{Introduction}\label{sec1}
\begin{figure*}[hbtp]
 \begin{center}
    \includegraphics[width = 13.5cm]{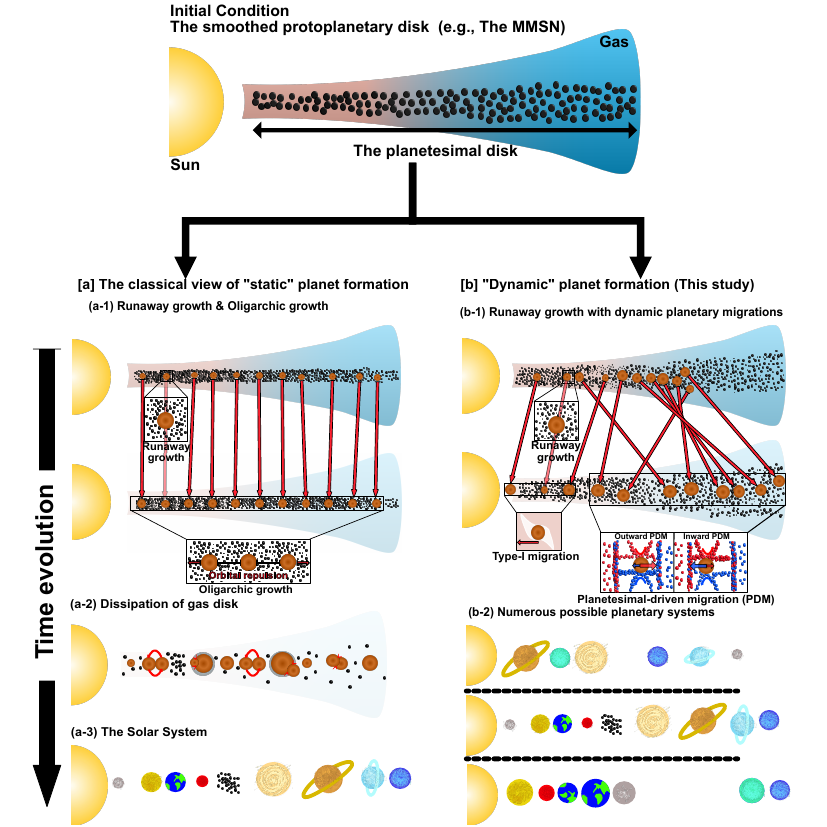}
 \end{center}
 \caption{\textbf{Two different planet formation scenarios starting from the same smoothed protoplanetary disk model.} [a] The classic scenario illustrating the formation of the solar system \citep{1972epcf.book.....S,1981PThPS..70...35H,1985prpl.conf.1100H}: Runaway growth of planetesimals form protoplanets and the orbital repulsion between them results in their in-situ oligarchic growth in the disk (a--1). Subsequently, as the disk gas dissipates, orbital crossings, giant impacts, and gas accretion take place, eventually forming the solar system (a-2 \& a-3). [b] The dynamic planet formation scenario realizing the formation of diverse planetary systems (this study): Planetary embryos formed through runaway growth migrate within the disk via outward/inward PDM and Type-I migration (b-1). Eventually, diverse planetary systems will be formed through planetary migration (b-2).\\{Alt text: Two schematic panels comparing two different planet formation scenarios. Panel a shows the classical view of static planet formation through in-situ runaway and oligarchic growth. Panel b shows dynamic planet formation from this study, where migrating embryos form diverse planetary systems.}}
 \label{fig:schematic}
\end{figure*}

According to the classical theory of planet formation, planets form through the accretion of numerous small bodies, planetesimals \citep{1972epcf.book.....S,1981PThPS..70...35H,1985prpl.conf.1100H}. In this theory, planetary embryos form through the runaway growth of planetesimals, in which larger planetesimals grow faster \citep{1978Icar...35....1G,1989Icar...77..330W,1993Icar..106..210I,1996Icar..123..180K}. The planetary embryos thus formed continue to grow and eventually evolve into terrestrial planets or the cores of gas and ice giants, depending on the place they formed (figure \ref{fig:schematic}a). This ``in-situ runaway and oligarchic growth model'' has long served as a starting point for planet formation theory, even though more recent models also emphasize additional channels, including pebble accretion (e.g., \cite{2012A&A...544A..32L,2014A&A...572A.107L}).

This in-situ model, however, faces several challenges in explaining key aspects of the planetary formation process in both our solar and exoplanetary systems. Notably, it is difficult to account for the formation of Uranus and Neptune within the solar system's lifetime \citep{2001Icar..153..224L,2002AJ....123.2862T}. Moreover, starting with the discovery of the exoplanet by \authorcite{1995Natur.378..355M} (\yearcite{1995Natur.378..355M}), more than 6,000 exoplanets, including hot and cold Jupiters or super-Earths, have been identified in recent years\footnote{The current number of discovered exoplanets can be found at ``Exoplanet Exploration: Planets Beyond Our Solar System.'' (\url{https://exoplanets.nasa.gov/})}. The diversity of exoplanetary systems is not easy to explain with the classical theory \citep{2010Sci...327..977B,2015JATIS...1a4003R}. The migration of (proto)planets, both inward and outward, seems necessary to explain the formation process of ice giants' cores and also diverse exoplanetary systems.

Indirect evidence of such migration is found in our solar system too; for instance, the existence of plutinos suggests that Neptune migrated outward during its formation process \citep{1993Natur.365..819M,1995AJ....110..420M,1999AJ....117.3041H}. Possible mechanisms for planetary migration include planet-gas disk interactions, such as Type-I migration, and planet-planetesimal interactions, known as planetesimal-driven migration (PDM) \citep{1986Icar...67..164W,2002ApJ...565.1257T,2000ApJ...534..428I,2009Icar..199..197K,2011Icar..211..819C,2014Icar..232..118M,2016ApJ...819...30K,2024PASJ...76.1309J}.

In \authorcite{2024PASJ...76.1309J} (\yearcite{2024PASJ...76.1309J}) (hereafter Paper I), we have conducted a large number of self-consistent $N$-body simulations of PDM, in which gravitational interactions among all planetesimals (self-stirring), the gas drag, and Type-I migration are all taken into account. We found that PDM can be effective even with the protoplanet-to-planetesimal mass ratio below 10 ($M_{\mathrm{proto}}/m_{\mathrm{planetesimal}}<10$), challenging the threshold of 100 by \authorcite{2014Icar..232..118M} (\yearcite{2014Icar..232..118M}) (see below). Moreover, we found that protoplanets can overcome inward drift caused by Type-I torque, with some undergoing outward migration via outward PDM. These results suggest that PDM serves as a rapid diffusion mechanism for planetary embryos, allowing for dynamic redistribution of protoplanets within the disk and significantly influencing the early stages of planet formation.

In \authorcite{2014Icar..232..118M} (\yearcite{2014Icar..232..118M}), it was argued that PDM should be monotonic to be effective.
In contrast, we proposed a different view in Paper I. We showed that planetary migration results from a combination of monotonic PDM and stochastic variations in migration direction caused by two-body relaxation. At lower mass ratios, the stochastic component becomes more important. These findings indicate that there is no single critical mass ratio threshold that determines the effectiveness of PDM. In Paper I, however, a single protoplanet embedded in the disk was adopted as the initial condition, limiting the applicability of their findings to more realistic scenarios in which multiple protoplanets undergo runaway and oligarchic growth (e.g., \cite{1998Icar..131..171K}). Thus, whether planets migrate through PDM within the disk and how such migration influences the planet formation process remain open questions.

In this paper, we report the results of the first self-consistent $N$-body simulations of planet formation from a large-scale planetesimal disk with sufficiently high resolution to investigate the effect of PDM, performed on the supercomputer Fugaku. The results of our simulations suggest that even within the classical protoplanetary disk framework, where we assume a minimum-mass solar nebula (MMSN) \citep{1981PThPS..70...35H} with a total mass enhanced by a factor of four ($4\times$MMSN), the planetary formation does not proceed ``in-situ'' as assumed in the standard model. Instead, planetary embryos migrate both inward and outward even in the early runaway growth phase (figure \ref{fig:schematic}b). Our results suggest that the basic assumption of the standard model and MMSN, that the planets are formed locally, is difficult to support. This is consistent with previous studies, which show that giant planets formed in the outer disk undergo substantial migration (\cite{2009ApJ...698..606C}) and that close-in super-Earths subject to eccentricity damping and type-I migration rapidly drift toward the inner disk edge rather than forming purely in-situ in MMSN-like disks (\cite{2015A&A...578A..36O}). Our simulations provide complementary evidence that planetary embryos likewise do not grow in-situ due to PDM and Type-I migration. Thus, dynamic planetary migration through PDM and Type-I migration play fundamental roles in shaping planetary systems.

Although recent theoretical and observational studies suggest that planetesimals and planetary embryos may form preferentially at
discrete radii where the solid surface density is locally enhanced,
producing ring-like substructures in protoplanetary disks
(e.g., \cite{2015ApJ...808L...3A,2022NatAs...6..357I,2023NatAs...7..330B,2024ApJ...972..181O}), in this study, we adopt the MMSN-like disk, which has long served as a representative baseline model in planet-formation studies. This choice is a first step toward investigating how PDM influences the planet formation process. In subsequent work, we will study planet formation in various disks.

This paper is organized as follows. In section \ref{method}, we introduce our simulation models and numerical methods. In section \ref{results}, we present our simulation results and discuss model parameter dependence. In section \ref{discussion_summary}, we briefly review the early $N$-body simulations of planet formation and discuss the implications of recent observations of protoplanetary disks and modern disk evolution models in light of our simulation results. We also discuss the potential impact of collisional fragmentation on PDM and conclude by summarizing our findings.

\section{Methods} \label{method}
\subsection{Disk model} \label{methods:disk}
Following MMSN \citep{1981PThPS..70...35H}, we define the solid surface density $\Sigma_{\mathrm{dust}}$, the gas surface density $\Sigma_{\mathrm{gas}}$, and gas disk temperature $T$ as functions of the radial heliocentric distance $r$ as follows:
\begin{align}
\Sigma_{\mathrm{dust}}&=10f_{\mathrm{dust}}\eta_{\mathrm{ice}}\left(\frac{r}{1~\mathrm{au}}\right)^{p}~\mathrm{g~cm}^{-2}, \label{eq:sigma_dust}\\
\Sigma_{\mathrm{gas}}&=2400f_{\mathrm{gas}}\left(\frac{r}{1~\mathrm{au}}\right)^{p}\exp\left(-\frac{t}{1~\mathrm{Myr}}\right)~\mathrm{g~cm}^{-2},\label{eq:sigma_gas}\\
T&=2.8\times10^2\left(\frac{r}{1~\mathrm{au}}\right)^{\beta}~\mathrm{K},\label{eq:temp}
\end{align}
where $f_{\mathrm{gas}}$ and $f_{\mathrm{dust}}$ represent scaling factors, while $\eta_{\mathrm{ice}}$ in equation (\ref{eq:sigma_dust}) denotes the enhancement due to ice condensation. Inside the snowline, we set $\eta_{\mathrm{ice}}=1$ while, $\eta_{\mathrm{ice}}=4.2$ beyond the snowline. In this study, taking into account the stellar evolution of the Sun and the viscous accretion of the gas disk, we assume the snowline to be at 2 au, as suggested by \authorcite{2011ApJ...738..141O} (\yearcite{2011ApJ...738..141O}) and \authorcite{2016ApJ...819...30K} (\yearcite{2016ApJ...819...30K}). In equations (\ref{eq:sigma_dust}) through (\ref{eq:temp}), the radial power indices $p$ and $\beta$ are set to $-3/2$ and $-1/2$, respectively, following \authorcite{1981PThPS..70...35H} (\yearcite{1981PThPS..70...35H}). To be consistent with Paper I, we model the gas dissipation as an exponential decay with a timescale of 1 Myr.

In our simulations, we use the value of $\Sigma_{\mathrm{dust}}$ four times that of MMSN, following previous studies \citep{2014Icar..232..118M,2024PASJ...76.1309J}. Thus, we set the value of the dust scaling factor $f_{\mathrm{dust}} = 2.84$, while the gas scaling factor $f_{\mathrm{gas}} = 0.71$. This indicates that the gas-to-dust ratio $f_{\mathrm{g/d}}$ in this study, expressed in terms of MMSN value denoted as $f_{\mathrm{MMSN}}$, is calculated as $f_{\mathrm{g/d}} = 0.25\times f_{\mathrm{MMSN}}\approx15$. Although our disk structure is dust-rich compared to MMSN, recent ALMA surveys show that protoplanetary disks exhibit a wide range of gas-to-dust mass ratios. The Molecules with ALMA at Planet-forming Scales (MAPS) ALMA Large Program (e.g., \cite{2021ApJS..257....1O}) inferred $f_{\mathrm{g/d}}\sim 10$--$100$ for its sample of nearby disks. In addition, the more recent AGE-PRO survey of 30 disks \citep{2025ApJ...989....1Z} found median gas-to-dust mass ratios of $\simeq 122$, $46$, and $120$ in Ophiuchus, Lupus, and Upper Sco, respectively. Their results indicate that typical disks have gas-to-dust ratios of order the interstellar medium ($\sim 100$), while some disks, especially in Lupus, show systematically lower values. Our adopted value $f_{\mathrm{g/d}}\approx 15$ is therefore lower than the typical median ratios inferred from large samples and should be regarded as representing a dust-rich disk at the lower end of the observationally inferred distribution.

\subsection{Gas drag model}\label{methods:gas}
The force formula of gas drag is given by \authorcite{1976PThPh..56.1756A} (\yearcite{1976PThPh..56.1756A}) as
\begin{equation}
\boldsymbol{F}_{\mathrm{drag}}=-\frac{1}{2m_{\mathrm{p}}}C_{\mathrm{D}}\pi r_{\mathrm{p}}^2\rho_{\mathrm{gas}}|\Delta \boldsymbol{v}|\Delta\boldsymbol{v}.\label{eq:gas_drag}
\end{equation}
Here, $C_{\mathrm{D}}$, $\rho_{\mathrm{gas}}$ and $\Delta \boldsymbol{v}$ represent the gas drag coefficient, the gas density and the relative velocity of the planetesimal to the disk gas, while $m_{\mathrm{p}}$ and $r_{\mathrm{p}}$ denote the mass and radius of the planetesimal, respectively. According to \authorcite{1976PThPh..56.1756A} (\yearcite{1976PThPh..56.1756A}), the gas drag coefficient $C_{\mathrm{D}}$ becomes approximately unity for the range of planetesimal masses considered in this study. Hence, we set $C_{\mathrm{D}}=1$. The gas density $\rho_{\mathrm{gas}}$ is given by
\begin{equation}
\rho_{\mathrm{gas}}=1.4\times10^{-9}f_{\mathrm{gas}}\left(\frac{r}{1~\mathrm{au}}\right)^{\alpha}~\mathrm{g~cm}^{-3},\label{eq:rho_gas}
\end{equation}
where $\alpha$ is the radial dependence and is set to $-11/4$. To calculate the relative velocity between the planetesimal and the disk gas $\Delta \boldsymbol{v}$, we need to estimate the disk gas velocity. The circular velocity of the disk gas is given by $v_{\mathrm{gas}}=v_{\mathrm{K}}(1-|\eta|)$ \citep{1976PThPh..56.1756A}. Here, $v_{\mathrm{K}}$ represents the Keplerian velocity, and $\eta$ is a dimensionless quantity that characterizes the pressure gradient of the gas disk and is calculated as follows: 
\begin{equation}
\eta=-\frac{1}{2}\frac{c_{\mathrm{s}}^2}{r^2\Omega^2}\left\lbrack\frac{\partial \log(\rho_{\mathrm{gas}}T)}{\partial \log r} \right\rbrack, \label{eq:eta}
\end{equation}
where $c_{\mathrm{s}}$ and $\Omega$ are the sound speed and the Keplerian angular velocity. These are given by
\begin{equation}
c_{\mathrm{s}}=\left(\frac{k_{\mathrm{B}}T}{\mu m_{\mathrm{H}}}\right)^{1/2}=1.0\times10^5\left(\frac{T}{280\mathrm~{K}}\right)^{1/2}~\mathrm{cm~s}^{-1}, \label{eq:sound}
\end{equation}
\begin{equation}
\Omega=\left(\frac{GM_{*}}{r^3}\right)^{1/2}=2.0\times10^{-7}\left(\frac{r}{\mathrm{au}}\right)^{-3/2} ~~\mathrm{s}^{-1}, \label{eq:angular}
\end{equation}
where $k_{\mathrm{B}}$, $\mu$, $m_{\mathrm{H}}$, $G$ and $M_{*}$ are the Boltzmann constant, the mean molecular weight of the gas, the mass of a hydrogen atom, the gravitational constant and the mass of the central star. The values of the mean molecular weight of the gas $\mu$ and the mass of the central star are set to 2.34 and 1$M_{\odot}$.

The radius of the planetesimal $r_{\mathrm{p}}$ is given as follows:
\begin{equation}
r_{\mathrm{p}}=\left(\frac{3m_\mathrm{p}}{4\pi \rho_{\mathrm{p}}}\right)^{1/3}, \label{eq:radius}
\end{equation}
where $\rho_{\mathrm{p}}$ is the internal density of the planetesimals, and we set the value of $\rho_{\mathrm{p}}$ to $2~\mathrm{gcm}^{-3}$. The initial masses of the planetesimals are set between $2.4 \times 10^{-4} M_{\oplus}$ and $4.9 \times 10^{-4} M_{\oplus}$, depending on the initial number of particles. Using equation (\ref{eq:radius}), these initial masses correspond to radii $r_{\mathrm{p}}$ between 500 km and 700 km, so that the size of the planetesimals in this study is larger by a factor up to two orders of magnitude compared to those estimated in various earlier studies (e.g., \cite{1973ApJ...183.1051G,2009Icar..204..558M,2015SciA....1E0109J}). 
Hence, the heating due to the gravitational interaction between planetesimals is enhanced, and the cooling effect of gas drag is reduced. In order to compensate for such an artificial effect, we use the following equation as gas drag instead of equation (\ref{eq:gas_drag}) in our simulation:
\begin{equation}
\boldsymbol{F}_{\mathrm{drag}}=-\frac{1}{2m_{\mathrm{p}}}C_{\mathrm{D}}\pi r_{\mathrm{p}}^2\mathcal{F}f_{\mathrm{MMSN}}\rho_{\mathrm{dust}}|\Delta\boldsymbol{v}|\Delta{\boldsymbol{v}} \label{eq:modified_gas_drag}
\end{equation}
Here $\mathcal{F}$ is introduced as an enhancement parameter for gas density, while $\rho_{\mathrm{dust}}$ represents the solid density, defined by $\rho_{\mathrm{dust}}=\rho_{\mathrm{gas}}/f_{\mathrm{g/d}}=4\rho_{\mathrm{gas}}/f_{\mathrm{MMSN}}$. We change the parameter from 50 to 1000 times the MMSN value ($\mathcal{F}$=[12.5, 25, 50, 100, 200, 250]). The range corresponds to that of the size of planetesimals from several kilometers to a hundred kilometers in radius (see also appendix \ref{appendix})\footnote{Similar approaches have been employed in earlier studies with $N$-body simulations (e.g., \cite{2011Natur.475..206W,2011Icar..211..819C,2015A&A...578A..36O}). Here we further extend our discussion to include the validity of these approaches (see appendix \ref{appendix}).}.

\subsection{Type-I migration model} \label{method:typeI}
Protoplanets that undergo runaway growth within the planetesimal disk begin to gravitationally perturb the disk gas and excite density waves. This damps orbital semi-major axis, eccentricity, and inclination of protoplanets \citep{1986Icar...67..164W,2002ApJ...565.1257T}. This damping of the orbital semi-major axis is generally called ``Type-I migration.''

We use the Type-I migration model developed by \authorcite{2020MNRAS.494.5666I} (\yearcite{2020MNRAS.494.5666I}). It is given by
\begin{equation}
\frac{\mathrm{d}\boldsymbol{v}}{\mathrm{d}t}=-\frac{v_{\mathrm{K}}}{2\tau_{a}}\boldsymbol{\mathrm{e}}_{\theta}-\frac{v_{r}}{\tau_{e}}\boldsymbol{\mathrm{e}}_{r}-\frac{v_{\theta}-v_{\mathrm{K}}}{\tau_e}\boldsymbol{\mathrm{e}}_{\theta}-\frac{v_z}{\tau_i}\boldsymbol{\mathrm{e}}_{z}, \label{eq:typeI}
\end{equation}
where $\boldsymbol{v}$, $\boldsymbol{\mathrm{e}}_{r}$, $\boldsymbol{\mathrm{e}}_{\theta}$, $\boldsymbol{\mathrm{e}}_{z}$ $\tau_{a}$, $\tau_{e}$ and $\tau_{i}$ are the planetary velocity, the unit vectors in radial, azimuthal and vertical directions, the evolution timescales for the semi-major axis, eccentricity, and inclination, respectively. These evolution timescales are given in Appendix D of \authorcite{2020MNRAS.494.5666I} (\yearcite{2020MNRAS.494.5666I}) and detailed in section 2.3 of Paper I:
\begin{eqnarray}
\tau_{a}^{-1}&\simeq C_{\mathrm{T}}h^2\left\lbrack1+\frac{C_{\mathrm{T}}}{C_{\mathrm{M}}}(\hat{e}^2+\hat{i}^2)^{1/2}\right\rbrack^{-1} t_{\mathrm{wave}}^{-1}, \label{eq:tau_a}\\
\tau_{e}^{-1}&\simeq0.780\left\lbrack1+\frac{1}{15}(\hat{e}^2+\hat{i}^2)^{3/2}\right\rbrack^{-1}t_{\mathrm{wave}}^{-1}, \label{eq:tau_e}\\
\tau_{i}^{-1}&\simeq0.544\left\lbrack1+\frac{1}{21.5}(\hat{e}^2+\hat{i}^2)^{3/2}\right\rbrack^{-1}t_{\mathrm{wave}}^{-1}.
\label{eq:tau_i}
\end{eqnarray}
Here, $C_{\mathrm{T}}=2.73-1.08p-0.87\beta$ and $C_{\mathrm{M}}=-6(2p-\beta-2)$ are the constant values where $p$ and $\beta$ are given in equations (\ref{eq:sigma_dust}), (\ref{eq:sigma_gas}) and (\ref{eq:temp}). Moreover, $\hat{e}$ and $\hat{i}$ are the eccentricity and inclination of the protoplanet, respectively, scaled by $h$. The characteristic timescale $t_{\mathrm{wave}}$ is given by \authorcite{2002ApJ...565.1257T} (\yearcite{2002ApJ...565.1257T}) as
\begin{equation}
t_{\mathrm{wave}}^{-1}=\left(\frac{M}{M_{*}}\right)\left(\frac{\Sigma_{\mathrm{gas}} a^2}{M_*}\right)h^{-4}\Omega,\label{eq:t_wave}
\end{equation}
where the aspect ratio $h$ of the gas disk at the semi-major axis $a$ of the protoplanet is given as 
\begin{equation}
h=\frac{H}{a}=3.3\times10^{-2}\left(\frac{a}{1\mathrm{au}}\right)^{(\beta+1)/2}. \label{eq:scale}
\end{equation}
Here, $H\equiv c_{\mathrm{s}}/\Omega$ is the scale height of the disk. 

\begin{table*}[h]
\tbl{List of models}{
\begin{tabular}{@{}lccccccc@{}}
\hline
Model & $N_{\mathrm{p,init}}$ & $m_{\mathrm{p}}$ ($M_{\oplus}$) & $\mathcal{F}$& $r_{\mathrm{d,in}}$ (au) & $r_{\mathrm{d,out}}$ (au) & $T_{\mathrm{end}}$ (Myr) & $N_{\mathrm{run}}$\\
\hline
1-1 & 237,520    & $4.9\times10^{-4}$   & $12.5$  & $2$ & $12$ & $4$ & $1$\\
1-2 & 237,520    & $4.9\times10^{-4}$   & $25$ & $2$ & $12$ & $4$ & $1$\\
1-3 & 237,520    & $4.9\times10^{-4}$   &  $50$ & $2$ & $12$ & $4$ & $1$\\
1-4 & 237,520    & $4.9\times10^{-4}$   & $100$  & $2$  & $12$ & $2.5$ & $1$\\
1-5 (Fiducial) & 237,520    & $4.9\times10^{-4}$   & $200$  & $2$ & $12$ & $2.5$ & $1$\\
1-6 & 237,520    & $4.9\times10^{-4}$   & $250$  & $2$ & $12$ & $2.5$ & $1$\\
2-1 & 354,350    & $4.9\times10^{-4}$   & $250$  & $2$ & $20$ & $1.2$ & $2$\\
2-2 & 708,700    & $2.45\times10^{-4}$   & $200$  & $2$ & $20$ & $1.2$ & $1$\\
\hline
\end{tabular}}\label{tab1}%
\end{table*}

\subsection{Numerical methods} \label{method:nemerical}
We use the $N$-body simulation code GPLUM \citep{2021PASJ...73..660I}. This code is characterized by its use of the Framework for Developing Particle Simulator (FDPS) \citep{2016PASJ...68...54I,2018PASJ...70...70N}, the particle-particle particle-tree (P$^3$T) scheme \citep{2011PASJ...63..881O}, and the individual cut-off method \citep{2021PASJ...73..660I}, all of which contribute to its exceptional scalability in parallel computing environments. Additionally, GPLUM is optimized for the supercomputer Fugaku \citep{2024PASJ...76.1309J}. Using GPLUM on Fugaku allows for simulations that consider gravitational interactions among all particles in large-scale planetesimal disks ($\sim$10 au) with a large number of particles ($N\gtrsim 10^{5}$). Here we summarize the features of GPLUM and describe how the modified gas drag model (equation \ref{eq:modified_gas_drag}) and the Type-I migration model (equation \ref{eq:typeI}) presented in sections \ref{methods:gas} and \ref{method:typeI} are implemented in the code.

GPLUM uses FDPS, a general-purpose, high-performance library designed for large-scale particle simulations. The concept of FDPS is to alleviate the complexity of large-scale parallelization tasks, such as domain decomposition, particle distribution, and particle data exchange, so that users can develop programs without focusing on the details of parallelization. Indeed, GPLUM has achieved exceptional scalability in parallel computing environments by using FDPS \citep{2021PASJ...73..660I}.

GPLUM uses the P$^3$T scheme \citep{2011PASJ...63..881O} which is a hybrid integrator based on Hamiltonian splitting (e.g., \cite{1999MNRAS.304..793C}). This scheme accelerates the gravitational calculations in $N$-body simulations by combining computational techniques from both non-collisional and collisional systems. Specifically, the integration of the gravitational interactions between particles is divided into hard and soft parts, based on the cutoff radius between each pair of particles. They are given by
\begin{align}
H&=H_{\mathrm{Hard}}+H_{\mathrm{Soft}}, \label{Hamiltonian}\\
H_{\mathrm{Hard}}&=\sum_{i}\left\lbrack\frac{|\boldsymbol{p}_i|^2}{2m_i}-\frac{GM_*m_i}{r_i}\right\rbrack\nonumber\\
&-\sum_{i}\sum_{j>i}\frac{Gm_im_j}{r_{ij}}[1-W(r_{ij};r_{\mathrm{out},ij})], \label{Hard_part}\\
H_{\mathrm{Soft}}&=-\sum_{i}\sum_{j>i}\frac{Gm_im_j}{r_{ij}}W(r_{ij};r_{\mathrm{out},ij}), \label{Soft_part}\\
r_{ij}&=|\boldsymbol{r_{i}}-\boldsymbol{r}_j|, 
\end{align}
where $m_{i}$,$~\boldsymbol{p}_{i}$, $\boldsymbol{r}_{i}$, $r_{\mathrm{out},ij}$, $W(r_{ij};r_{\mathrm{out},ij})$, and $r_{ij}$ are the mass, momentum, position of the $i$th particle, the cutoff radius, the cutoff function for the Hamiltonian and the distance between $i$th and $j$th particles. The cutoff radius between the $i$th and $j$th particles $r_{\mathrm{out},ij}$ is given by:
\begin{eqnarray}
r_{\mathrm{out},ij}&=&\max(\tilde{R}_{\mathrm{cut},0}r_{\mathrm{Hill},i},\tilde{R}_{\mathrm{cut},1}v_{\mathrm{ran},i}\Delta t,\nonumber\\&&
\tilde{R}_{\mathrm{cut},0}r_{\mathrm{Hill},j},\tilde{R}_{\mathrm{cut},1}v_{\mathrm{ran},j}\Delta t), \label{eq:cut-off_radii}
\end{eqnarray}
where $\tilde{R}_{\mathrm{cut},0}=2$ and $\tilde{R}_{\mathrm{cut},1}=4$ are the dimensionless parameters. The terms $r_{\mathrm{Hill},i}$ and $r_{\mathrm{Hill},j}$ are the Hill radii of the $i$th and $j$th particles, while $v_{\mathrm{ran},i}$ and $v_{\mathrm{ran},j}$ are the root mean square random velocities of particles around the $i$th and $j$th particles, respectively. Here, the random velocity is defined as the difference between the particle's velocity and the Keplerian velocity. The Hill radius is given by
\begin{equation}
r_{\mathrm{Hill,}i}=\left(\frac{m_{i}}{3M_*}\right)^{1/3}a_{i}, \label{eq:Hill_radius}
\end{equation}
where $a_i$ is the semi-major axis of the $i$th particle. The cutoff function $W(r_{ij};r_{\mathrm{out},ij})$ has the following form \citep{2017PASJ...69...81I}:
\begin{equation}
W(y;\gamma)=\left\{
\begin{array}{ll}
\frac{7\gamma^6-9\gamma^5+45\gamma^4-60\gamma^3\ln\gamma-45\gamma^2+9\gamma-1}{3(\gamma-1)^7}y, &\\
    \hspace{4.0cm} (y<\gamma),&\\
    f(y;\gamma)+[1-f(1;\gamma)y], &\\
    \hspace{3.5cm}(\gamma\leq y<1),&\\
        1, &\\
       \hspace{4.0cm} (1\leq y), &\\
\end{array}\right. \label{eq:cutoff_fuction}
\end{equation}
where
\begin{align}
f(y;\gamma)=&\lbrace-10/3y^7+14(\gamma+1)y^6\nonumber\\
            &-21(\gamma^2+3\gamma+1)y^5\nonumber\\
            &+\lbrack 35(\gamma^3+9\gamma^2+9\gamma+1)/3\rbrack y^4\nonumber\\
            &-70(\gamma^3+3\gamma^2+\gamma)y^3\nonumber\\
            &+210(\gamma^3+\gamma^2)y^2-140\gamma^3y\ln y\nonumber\\
            &+(\gamma^7-7\gamma^6+21\gamma^5-35\gamma^4)\rbrace/(\gamma-1)^7.
\end{align}
Here $y=r_{ij}/r_{\mathrm{out},ij}$ and $\gamma$ is a constant parameter in the range $0$ to $1$, which we set to $\gamma=0.5$. Equation (\ref{eq:cutoff_fuction}) indicates that when the distance $r_{ij}$ exceeds their cutoff radius ($r_{ij} > r_{\mathrm{out},ij}$), it becomes unity. Thus, gravitational forces in the hard part affect only particles that are closer than $r_{\mathrm{out},ij}$. Based on equation (\ref{eq:cut-off_radii}), interactions between particles within the cutoff radius are classified as the hard part and are integrated using the fourth-order Hermite scheme \citep{1991ApJ...369..200M} with the block individual time step scheme \citep{1963MNRAS.126..223A,1986LNP...267..156M}. Interactions outside the cutoff radius are classified as the soft part and are integrated using the Barnes-Hut tree scheme \citep{1986Natur.324..446B} with a constant time step, available in FDPS.

The cutoff radius defined in equation (\ref{eq:cut-off_radii}) is set individually for each particle pair. This approach is known as the individual cut-off method \citep{2021PASJ...73..660I}. Prior to GPLUM, P$^3$T implementations such as PENTACLE \citep{2017PASJ...69...81I} used a shared cut-off, where the cutoff radius was determined based on the Hill radius of the most massive particle in all interactions. 
However, the shared cut-off method commonly results in an overestimation of cutoff radii for many particle pairs, thereby decreasing the efficiency of Hamiltonian splitting \citep{2021PASJ...73..660I}. To solve this problem, GPLUM implements the individual cut-off method to overcome the decrease in the efficiency of Hamiltonian splitting.

We modified the integration step in GPLUM to include the effects of the modified gas drag and the Type-I torque given by equations (\ref{eq:modified_gas_drag}) and (\ref{eq:typeI}):
\begin{equation}
\mathcal{G}^{\Delta t/2}\mathcal{K}^{\Delta t/2}\mathcal{D}^{\Delta t}\mathcal{K}^{\Delta t/2}\mathcal{G}^{\Delta t/2},
\end{equation}
where $\Delta t$ denotes the time step and each symbol represents an operator.
The operator $\mathcal{D}$ represents the time integration for each hard-part cluster, and $\mathcal{K}$ gives the velocity kick due to the soft part.
The operator $\mathcal{G}$ corresponds to the velocity kick caused by gas drag and the Type-I torque. As in the case of the soft part, it is applied twice, referred to as the first and second velocity kicks, each applied for half of the time step $\Delta t$ before and after the operator $\mathcal{D}$.

\subsection{Initial conditions} \label{method:initial}
We conducted a total of nine simulations. Table \ref{tab1} shows the names of models, the initial number of particles ($N_{\mathrm{p,init}}$), the initial mass of planetesimals ($m_{\mathrm{p}}$), the value of enhancement parameter for aerodynamic gas drag ($\mathcal{F}$), the inner and outer limits of the planetesimal disk ($r_{\mathrm{d,in}}$ and $r_{\mathrm{d,out}}$), the simulation end time ($T_{\mathrm{end}}$), and the number of runs ($N_{\mathrm{run}}$). As shown in table \ref{tab1}, all simulations are categorized into model 1 and model 2 depending on $r_{\mathrm{d,out}}$. Specifically, in model 1, $r_{\mathrm{d,in}}$ and $r_{\mathrm{d,out}}$ are set at 2 au and 12 au, respectively. In model 2, they are set at 2 au and 20 au, respectively. In all simulations, the gaseous disk extends from 0.1 to 20 au for model 1 and from 0.1 to 24 au for model 2, and these radii also define the inner and outer boundaries at which particles are removed from the simulation. The inner gas radius of 0.1 au is chosen because of computational resource limitations, so as to keep the time step for the soft part of the integration reasonably large in our large-scale simulations. The outer gas radius is set a few au beyond the outer edge of the initial planetesimal disk because outward migration driven by PDM becomes inefficient once the local planetesimal surface density drops significantly. 

Model 1 is further divided into models 1-1 through 1-6, based on the magnitude of gas drag (see also section \ref{methods:gas} in the Methods section). Model 2 is categorized into model 2-1 and model 2-2, based on the initial number of particles. Here, we note that we performed two runs with the same parameters but different random number seeds for model 2-1 to investigate the variation of the results due to the random number seed.

In all simulations, we used an axisymmetric surface density distribution as given by equation (\ref{eq:sigma_dust}). The initial eccentricities and inclinations of the planetesimals follow a Gaussian distribution with the dispersion $\langle e^2\rangle ^{1/2}=2\langle i^2\rangle ^{1/2}=2r_{\mathrm{Hill}}/a_{\mathrm{p}}$, where $r_{\mathrm{Hill}}$ is the Hill radius of the planetesimal given in equation (\ref{eq:Hill_radius}) with the semi-major axis $a_{\mathrm{p}}$ \citep{1992Icar...96..107I}. For the integration time step, a constant time step of $\Delta t=2^{-4}\mathrm{yr}/2\pi\approx10^{-2}\mathrm{yr}$ was used for the calculation of the soft part with Barnes-Hut tree scheme. For the hard part, where we employed the fourth-order Hermite scheme with block individual time steps, the minimum allowed time step was set to $\Delta t_{\mathrm{min}} = 2^{-30}\,\mathrm{yr}/2\pi \approx 1.5\times10^{-10}\,\mathrm{yr}$. We adopted an accuracy parameter $\eta_{\mathrm{H}}=0.02$ for the fourth-order Hermite scheme, 
which controls the individual time steps. All planetesimals were subject to gas drag and gravitational interactions (self-stirring). Additionally, we applied the Type-I torque to planetesimals that had grown to masses greater than $10^{-2} M_{\oplus}$.\footnote{This threshold was adopted because protoplanets beyond the snowline with masses exceeding $10^{-2} M_{\oplus}$ have a Type-I migration timescale of $\sim 10^6$ years, which is comparable to the planet formation timescale \citep{2002ApJ...565.1257T}.} Here, we note that perfect accretion was assumed for both planetesimal-planetesimal and planet-planetesimal collisions; hence, the effects of fragmentation were ignored in all simulations.

\section{Results}\label{results}
\subsection{Results of the fiducial model} \label{results:fiducial_model}
\begin{figure}[hbtp]
 \includegraphics[width= 8.0cm]{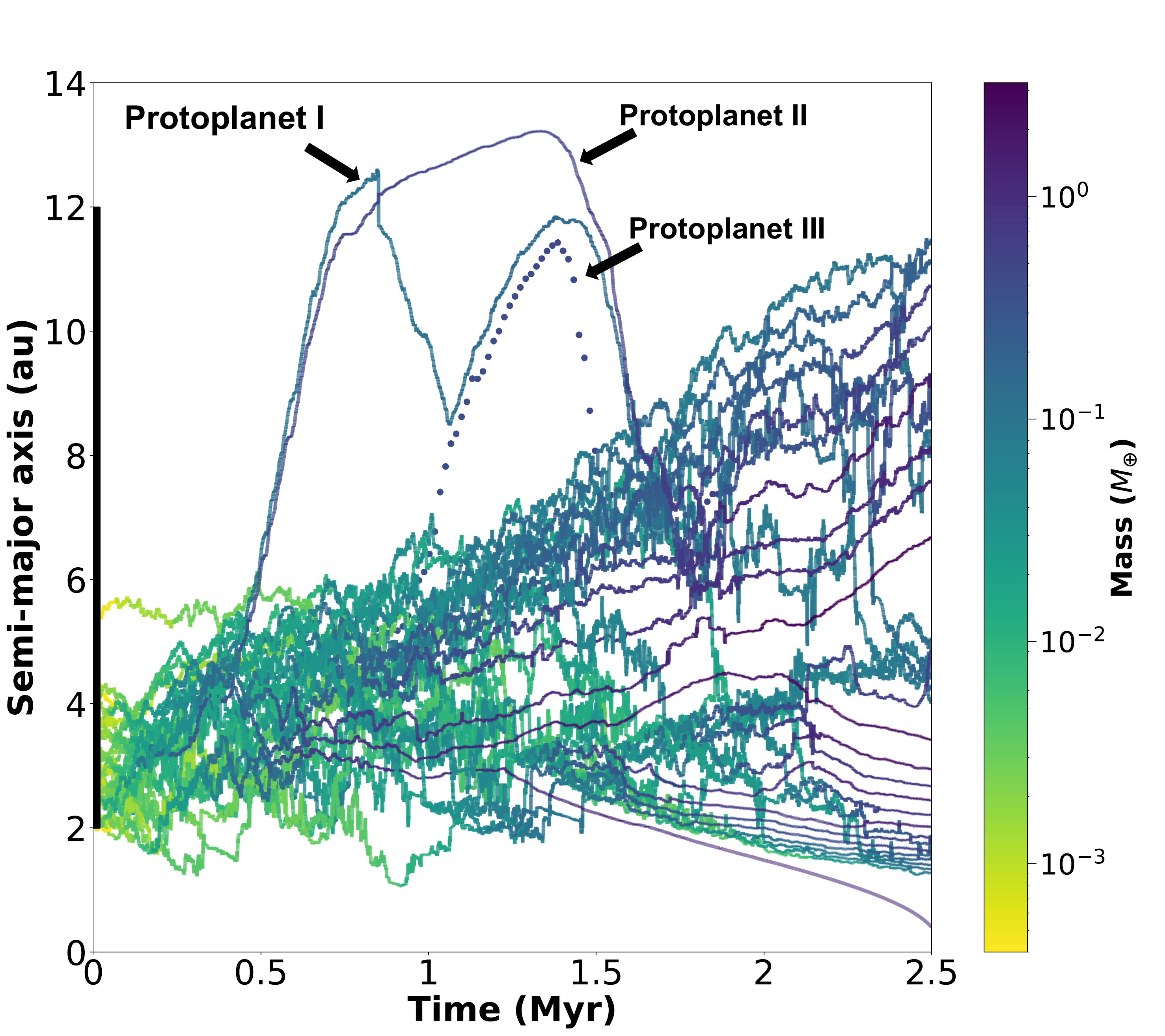}
 \caption{\textbf{The time evolution of the semi-major axes for the heaviest 30 bodies at 2.5 Myr} (including a body that collided with another body at approximately 1.87 Myr, indicated by a dotted curve). The color of each curve represents each mass. The black hatched regions on the left show the initial planetesimal disk size.\\{Alt text: The line graph showing the time evolution of semi-major axes and masses of planetary embryos formed in model 1-5. The horizontal axis represents time, and the vertical axis displays the semi-major axis in au.}}
 \label{fig:semimajor_grid}
\end{figure}

\begin{figure*}[hbtp]
 \begin{center}
    \includegraphics[width = 14.2cm]{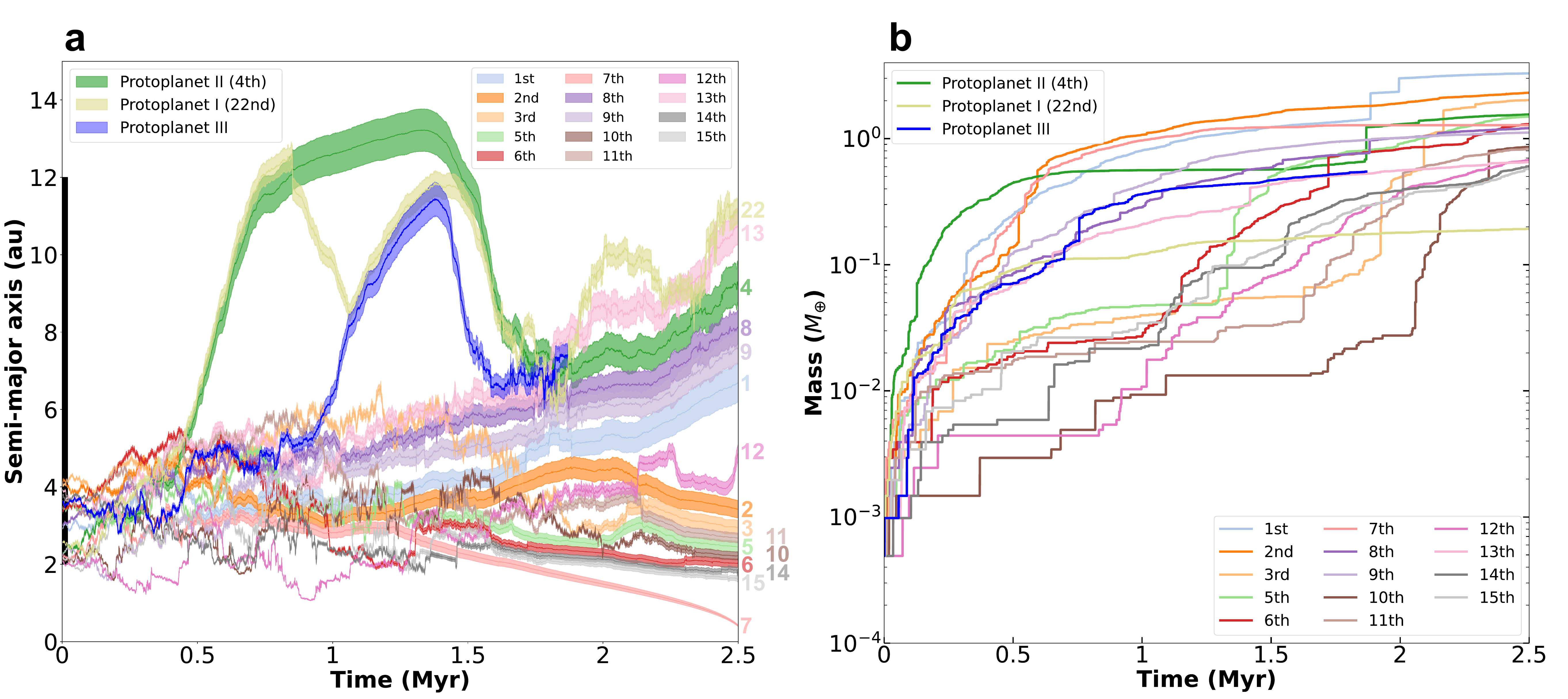}
 \end{center}
 \caption{\textbf{The time evolution of the semi-major axis and mass for the top 15 heaviest bodies at 2.5 Myr} (including protoplanet I (thin yellow curve), II (green curve), and III (blue curve), which collided with another body at approximately 1.87 Myr). (a) The width of each curve represents the orbital repulsion for each body, calculated as $a \pm 5r_{\mathrm{Hill}}$, where $a$ and $r_{\mathrm{Hill}}$ are the semi-major axis and the Hill radius. The black-hatched regions indicate the initial planetesimal disk size. The numbers on the right indicate the ranks based on the masses of the planetary embryos. (b) The color of each curve corresponds to the rank shown in (a).\\{Alt text: Two line graphs showing the orbital and mass evolution of the 15 heaviest protoplanets formed in model 1-5. Panel (a) shows the semi-major axis and orbital repulsion of planetary embryos, and panel (b) shows the color-coded mass ranking of the 15 heaviest protoplanets.}}
 \label{fig:semimajor}
\end{figure*}

Figures \ref{fig:semimajor_grid} and \ref{fig:semimajor} show the orbital semi-major axis and mass evolution of planetary embryos formed in our fiducial model (model 1-5). There are two remarkable features in these figures:
\begin{enumerate}
    \item The presence of three protoplanets that rapidly migrate outward around 0.5 Myr and 1 Myr, indicating that PDM can drive very fast outward migration. Although these protoplanets eventually fall back due to the truncated outer edge of the initial disk, it is expected that they would have continued migrating farther if the disk had extended outward.
    \item Even those that do not undergo such extreme migration exhibit a widespread distribution. By 2.5 Myr, planetary embryos have spread out to 12 au, although most of them initially started growing between 2 au and 4 au. 
\end{enumerate}
In the following, we discuss these two features in more detail.
%

Figure \ref{fig:semimajor_grid} shows that three planetary embryos undergo rapid outward migration through outward PDM. The two embryos (hereafter protoplanets I and II) broke away from their birth region (2-4 au) around 0.5 Myr and migrated outward, reaching the outer edge of the disk (12 au) by 0.73 and 0.83 Myr, respectively. Beyond the outer edge of the initial planetesimal disk, the migration naturally slowed down because the local reservoir of planetesimals was depleted and only planetesimals that had previously been scattered outward remained available for further encounters. Around 0.85 Myr, protoplanet I started to fall after an orbital crossing with protoplanet II driven by inward PDM and Type-I migration. Then, it encountered another embryo shown in dots (protoplanet III), which triggered another reversal and carried protoplanet I outward to 11.8 au by 1.35 Myr. Protoplanets I, II, and III subsequently began migrating inward through inward PDM but promptly encountered embryos migrating outward from the inner disk, leading to a reversal in their migration direction. By 2.5 Myr, protoplanets I and II are at 11 au and 9.2 au, respectively, while protoplanet III collided with another embryo at 7.4 au around 1.9 Myr. To quantify the relative importance of PDM and Type-I migration for these embryos, we estimate the characteristic migration timescales of both processes and follow the time evolution of the ratio $\tau_{\mathrm{TypeI}}/\tau_{\mathrm{PDM}}$ for protoplanets I-III (see appendix \ref{appendix2}). During the migration reversals of protoplanets I and II, $\tau_{\mathrm{TypeI}}/\tau_{\mathrm{PDM}}$ is of order $10^{1}-10^{2}$, indicating that PDM always overcomes Type-I migration in these phases.

Furthermore, figure \ref{fig:semimajor_grid} shows that embryos initially grown between 2 au and 6 au undergo rapid radial diffusion across the disk due to the combined effects of PDM, orbital repulsion, and Type-I migration. This behavior is consistently observed in all of our simulations (see also sections \ref{results:gas_drag} and \ref{results:disk_size}) and suggests a new paradigm of planet formation, in which protoplanets continue to grow while migrating through the disk, rather than stagnating at their isolated masses after depleting local feeding zones, as assumed in the classical theory \citep{2002ApJ...581..666K}.

Figure \ref{fig:semimajor}a shows the time evolution of the semi-major axes of protoplanets, with the width of each curve indicating the separation of 5 $r_{\mathrm{Hill}}$, which is the effective range of the orbital repulsion \citep{1995Icar..114..247K}. In figure \ref{fig:semimajor}a, planetary embryos that initially started growing between 2 au and 4 au have spread over a much wider radial range of 1 au to 12 au by 2.5 Myr. As shown in figure \ref{fig:semimajor}b, the masses of the largest and second-largest embryos reach 3 $M_{\oplus}$ and 2 $M_{\oplus}$, which are comparable to HD20794b and c \citep{2011A&A...534A..58P,2025A&A...693A.297N}, or protoplanets of gas and ice giants. The migration of the largest and second-largest embryos, through the combined effects of PDM and orbital repulsion, induced outward migration of embryos beyond the largest and inward migration of those interior to the second-largest. This resulted in a gap in the radial distribution of planetary embryos. At the end of the simulation, six protoplanets migrate outward via a combination of outward PDM and orbital repulsion (figure \ref{fig:semimajor}a); among them, the most massive one exceeds 2 $M_{\oplus}$, while even the smallest has a mass of approximately 0.2 $M_{\oplus}$ (figure \ref{fig:semimajor}b), indicating that each of them has sufficient mass to contribute to the formation of the rocky or icy cores of the ice giants. In addition to the fiducial model, gap structures are also observed in models 1-3, 1-6, 2-1 (Runs 1 and 2), and 2-2 (see sections \ref{results:gas_drag} and \ref{results:disk_size}). This result indicates that such a structure may commonly form during the process of planet formation.

\begin{figure*}[hbtp]
\begin{center}
 \includegraphics[width= 15.0cm]{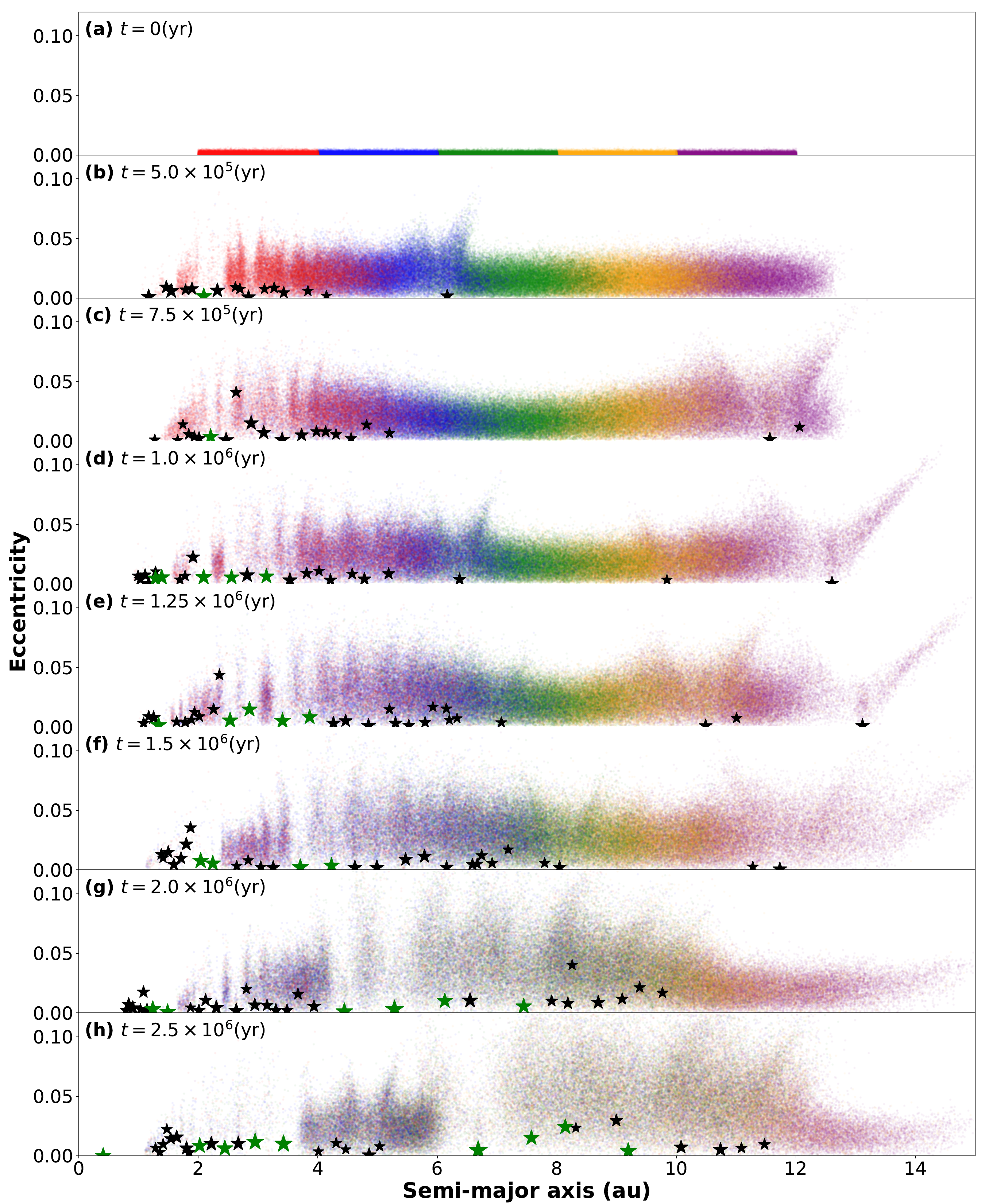}
\end{center}
 \caption{\textbf{The time evolution of the system during planet formation within the initially smoothed planetesimal disk.} Bodies with $0.1M_{\oplus}\leq M\leq M_{\oplus}$ and $M_{\oplus}\leq M$ are represented by black and green stars, respectively, scaled by $M^{1/3}$, where $M$ is the mass. The dots, colored red, blue, green, yellow, and purple, represent planetesimals. Each color indicates their initial semi-major axes: red for 2-4 au, blue for 4-6 au, green for 6-8 au, yellow for 8-10 au, and purple for 10-12 au. \textbf{(a)} Initial planetesimal disk with a surface density distribution of 4$\times$MMSN. \textbf{(b)} By 0.5 Myr, a single planetary embryo migrates outward to approximately 7 au. \textbf{(c)} By 0.75 Myr, two planetary embryos reach the outer edge of the disk ($\sim12$ au). \textbf{(d)} By 1 Myr, one of the two planetary embryos that reached the disk's outer edge changes its migration direction inward and reaches near 10 au, while a new embryo from the inner disk begins migrating outward. \textbf{(e)} By 1.25 Myr, the inward migrating planetary embryo encounters the outward migrating one, causing both to migrate outward. \textbf{(f)} By 1.5 Myrs, all planetary embryos that migrated outward begin migrating inward. \textbf{(g)} By 2 Myr, they encounter a group of outward migrating planetary embryos, \textbf{(h)} prompting them to begin migrating outwards again by 2.5 Myr.\\ {Alt text: Snapshots of model 1-5, arranged chronologically from panel (a) to panel (h). The horizontal axis represents semi-major axis in au, and the vertical axis displays eccentricity.}}
 \label{fig:total}
\end{figure*}
Figure \ref{fig:total} shows several snapshots for our fiducial model. The planetesimal disk is initially distributed smoothly at $t = 0$ yr (panel a). Within the disk, planetary embryos form through runaway growth. Some of these embryos migrate through PDM from an early runaway growth stage (panels b-f). By $t = 2$ Myr (panel g), the largest and second-largest embryos located near 5 au start migrating inward and outward due to orbital repulsion and subsequent PDM. They effectively push out neighboring embryos located both inside and outside their orbits, resulting in a bifurcation of the planetary distribution.

The results of our fiducial model indicate that, even under the classical disk framework of MMSN, the ``in-situ'' planet formation assumed in the canonical planet formation theory is unlikely to occur. Instead, protoplanets undergo substantial radial migration across the disk due to PDM and Type-I migration.

\subsection{Results of Model 1: Influence of initial planetesimal size} \label{results:gas_drag}
In all our simulations, the initial planetesimal radius is of the order of several hundred kilometers. This is larger by one to two orders of magnitude than the actual size of planetesimals estimated in earlier studies (e.g., \cite{1973ApJ...183.1051G,2009Icar..204..558M,2015SciA....1E0109J}). As a result, the heating due to gravitational interactions among planetesimals is enhanced, while the cooling effect of gas drag is reduced. In order to enhance the gas drag, we introduce $\mathcal{F}$ to the gas drag term.

Here we present the results of five simulations (models 1-1 to 1-4 and 1-6), in which the magnitude of gas drag $\mathcal{F}$ is treated as a free parameter in order to represent initial planetesimal sizes ranging from a few kilometers to several tens of kilometers in radius (see section \ref{methods:gas} and appendix \ref{appendix} for more details). In all simulations, the initial range of the planetesimal disk is set from $r_{\mathrm{d,in}} = 2$ au to $r_{\mathrm{d,out}} = 12$ au, and the disk is realized with 237,520 planetesimals, each having a radius of $r_{\mathrm{p}} \simeq 700$ km. Additionally, models 1-1 to 1-3 were simulated for 4 Myr, while models 1-4 and 1-6 were simulated for 2.5 Myr. Each model is summarized in table \ref{tab1}.

\begin{figure*}[hbtp]
 \begin{center}
 \includegraphics[width= 15.0cm]{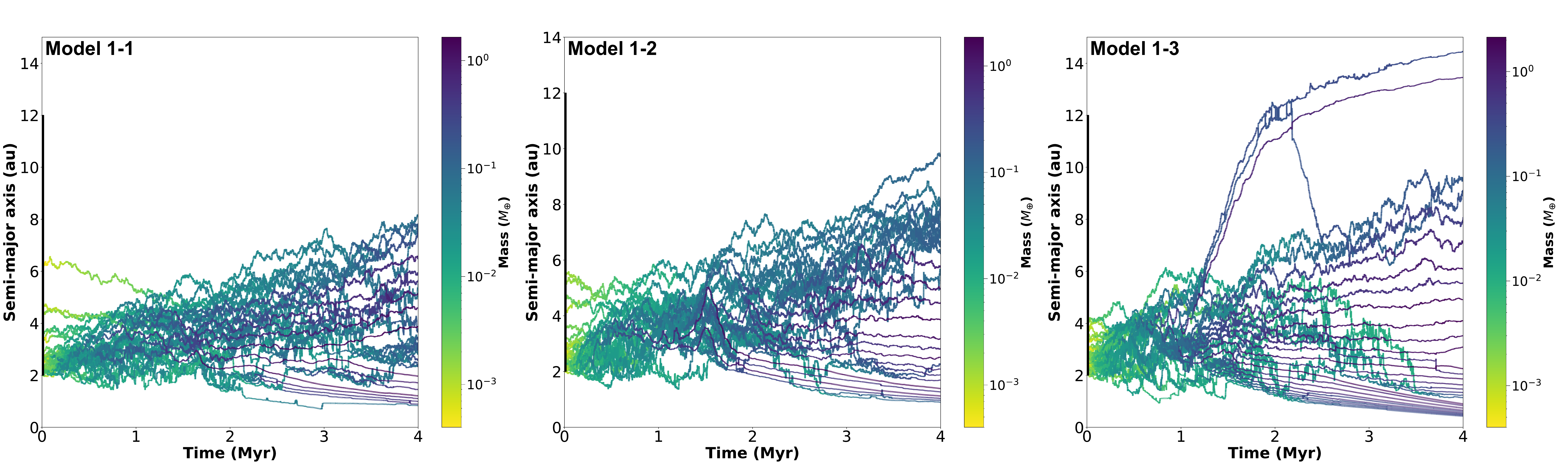}
 \end{center}
 \caption{\textbf{The time evolution of the semi-major axes for the heaviest 30 bodies at 4 Myr.} From left to right, the panels represent models 1-1 (left), 1-2 (middle), and 1-3 (right). The black hatched regions show the initial planetesimal disk size. \\{Alt text: The line graphs showing the time evolution of semi-major axes and masses of protoplanets formed in model 1-1, 1-2 and 1-3. The horizontal axis represents time, and the vertical axis displays the semi-major axis in au.}}
 \label{fig:semi-major1-1_3}
\end{figure*}

\begin{figure*}[hbtp]
 \begin{center}
 \includegraphics[width= 13.0cm]{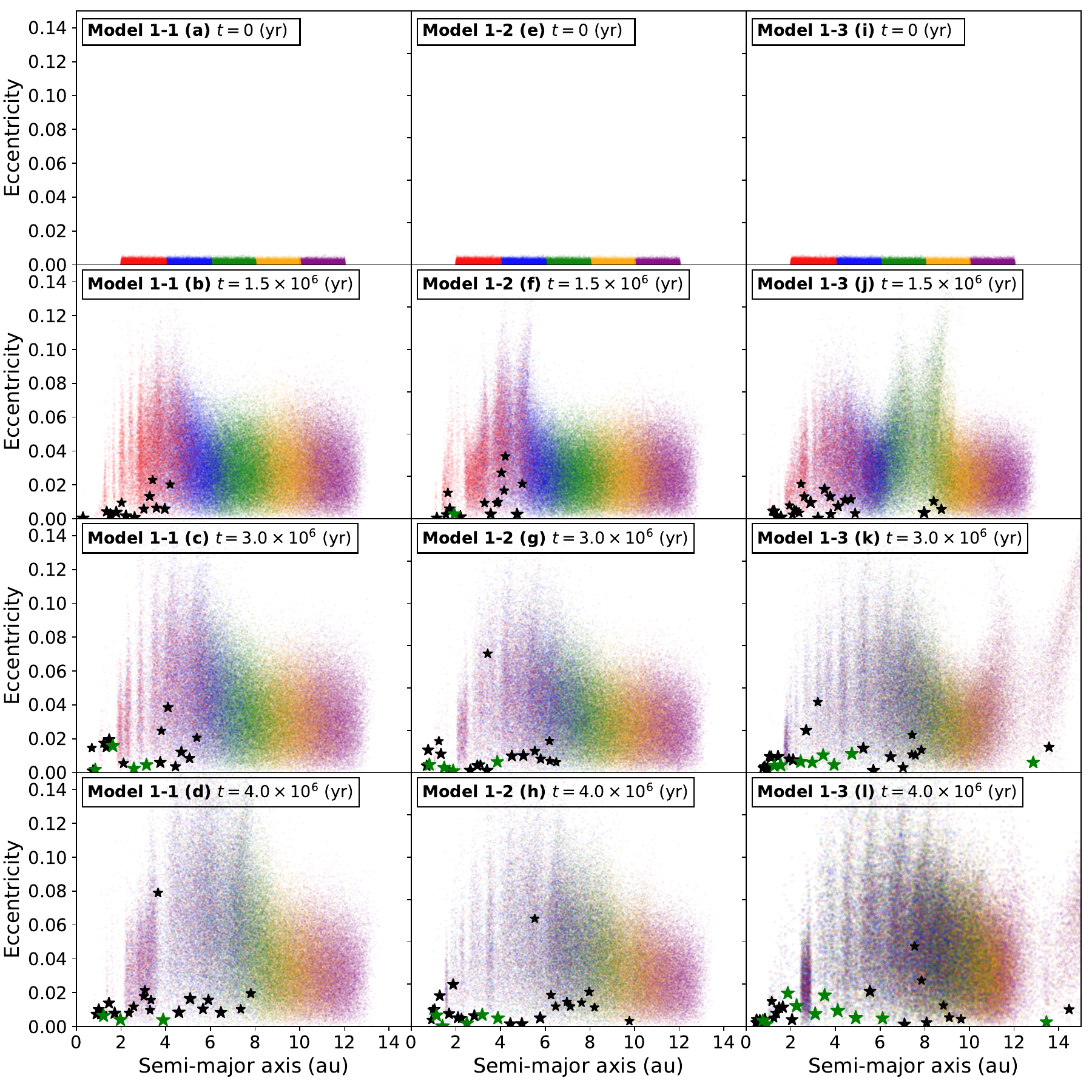}
 \end{center}
 \caption{\textbf{The time evolution of the systems during planet formation within the planetesimal disk.} From left to right, the panels represent models 1-1, 1-2, and 1-3 (see table 1 in the Methods section), and from top to bottom, they show the time evolution of the system in chronological order. Bodies with $0.1M_{\oplus}\leq M\leq M_{\oplus}$ and $M_{\oplus}\leq M$ are represented by black and green stars, respectively, scaled by $M^{1/3}$, where $M$ is the mass. The dots, colored red, blue, green, yellow, and purple, represent planetesimals. Each color indicates the planetesimals' initial semi-major axes: red for 2-4 au, blue for 4-6 au, green for 6-8 au, yellow for 8-10 au, and purple for 10-12 au.\\ {Alt text: Snapshots of model 1-1, 1-2, and 1-3, arranged chronologically from panel (a) to (d), (e) to (h) and (i) to (l), respectively. The horizontal axis represents semi-major axis in au, and the vertical axis displays eccentricity.}}
 \label{fig:total1}
\end{figure*}

\begin{figure*}[hbtp]
 \begin{center}
 \includegraphics[width= 14.0cm]{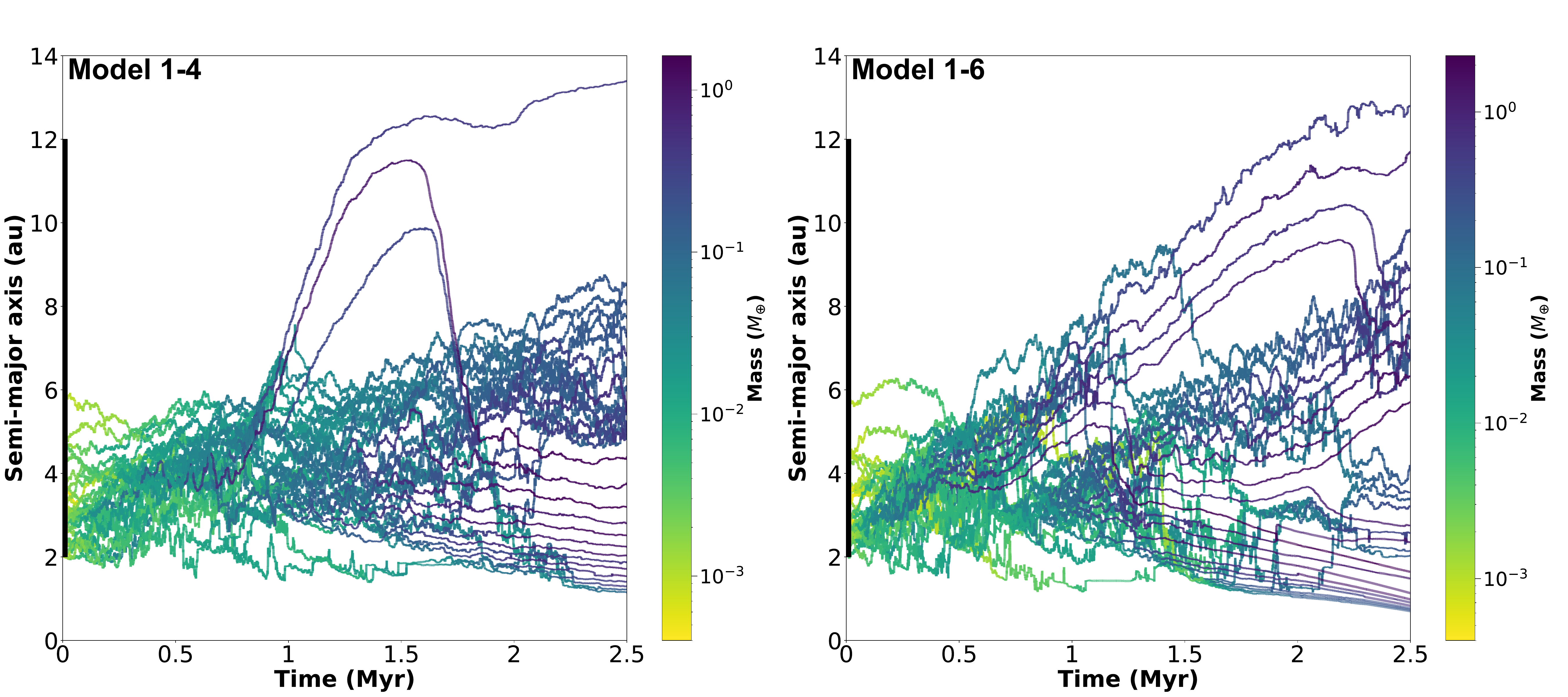}
 \end{center}
 \caption{\textbf{Same as figure \ref{fig:semi-major1-1_3}, but for models 1-4 (left) and 1-6 (right).} \\{Alt text: The line graphs showing the time evolution of semi-major axes and masses of protoplanets formed in model 1-4 and 1-6. The horizontal axis represents time, and the vertical axis displays the semi-major axis in au.}}
 \label{fig:semi-major1-4_6}
\end{figure*}

\begin{figure*}[hbtp]
 \begin{center}
 \includegraphics[width= 10.0cm]{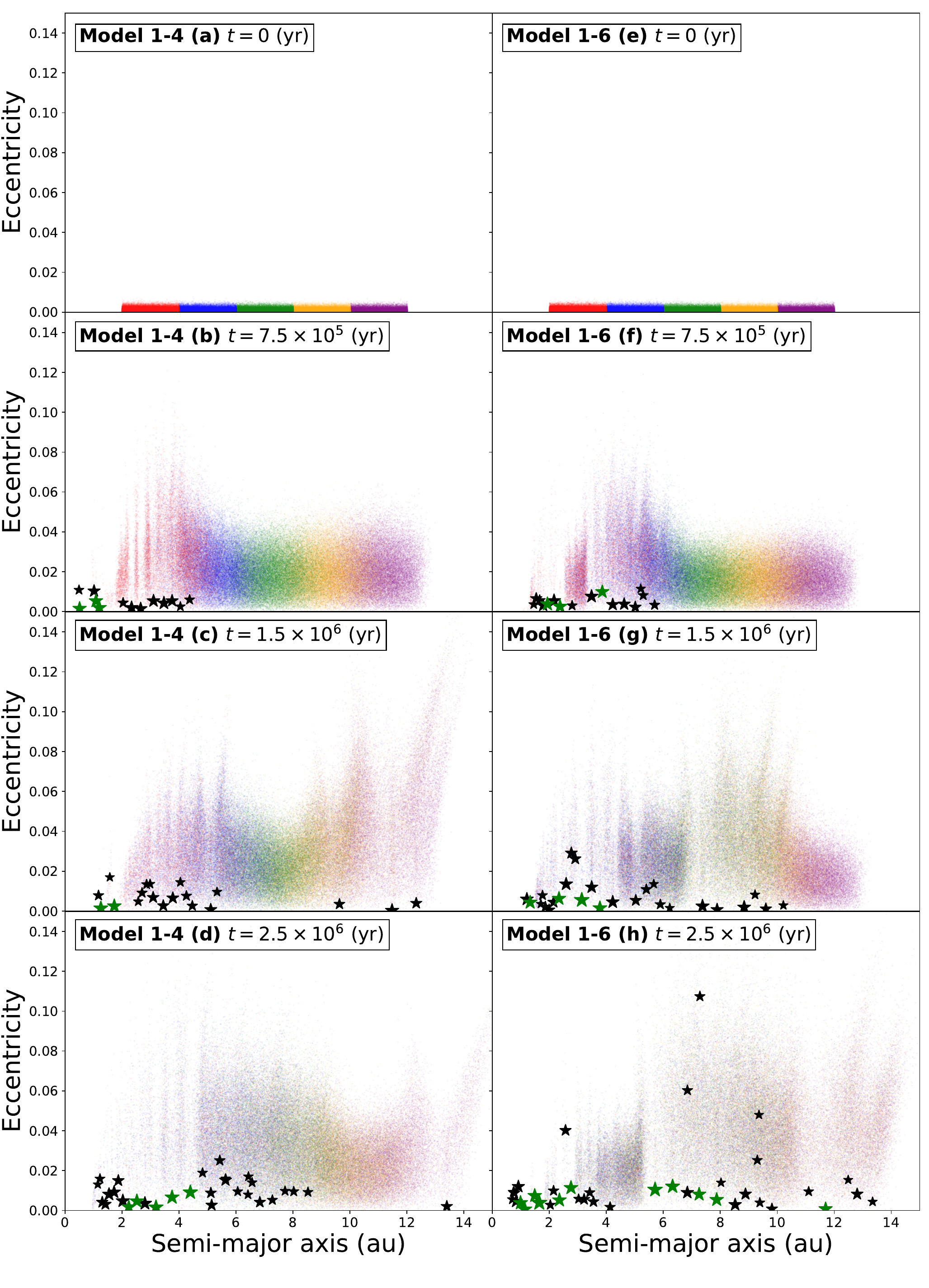}
 \end{center}
 \caption{\textbf{Same as figure \ref{fig:total1}, but for models 1-4 (left) and 1-6 (right).} \\ {Alt text: Snapshots of model 1-4 and 1-6, arranged chronologically from panel (a) to (d) and (e) to (h), respectively. The horizontal axis represents semi-major axis in au, and the vertical axis displays eccentricity.}}
 \label{fig:total1-2}
\end{figure*}

Figures \ref{fig:semi-major1-1_3} and \ref{fig:total1} show the orbital evolution of planetary embryos formed in models 1-1 to 1-3 and the snapshots. Likewise, figures \ref{fig:semi-major1-4_6} and \ref{fig:total1-2} show the corresponding results for models 1-4 and 1-6. In each column of figures \ref{fig:total1} and \ref{fig:total1-2}, snapshots of simulations are arranged in chronological order from top to bottom.

%

%

%

In models 1-1 to 1-3, we set the gas-density enhancement parameter $\mathcal{F}$ to $12.5$, $25$, and $50$ (see section~\ref{methods:gas} and appendix \ref{appendix}). These choices correspond to the effective planetesimal radius $r'_{\mathrm{p}}\simeq700/12.5=56~\mathrm{km}$ in model 1-1, $700/25=28~\mathrm{km}$ in model 1-2, and $700/50=14~\mathrm{km}$ in model 1--3. In all models, embryos form mainly between 2 and 5~au during runaway growth and then disperse through outward and inward PDM, Type-I migration, and orbital repulsion. In model 1-1, the thirty most massive bodies occupy about 0.85 to 8.1~au by 4~Myr, as seen in the left panels of figure \ref{fig:semi-major1-1_3} and panels a-d of figure \ref{fig:total1}. In model 1-2, planetary embryos extend to about 0.9 to 9.7~au, and the outermost embryo that formed in the inner disk migrates outward through PDM, as shown in the middle panel of figure \ref{fig:semi-major1-1_3} and panels e-h of figure \ref{fig:total1}. In model 1-3 (the right panels of figure \ref{fig:semi-major1-1_3} and panels i-l of figure \ref{fig:total1}), several embryos start rapid migration through outward PDM at about 1.2~Myr and reach the outer edge of the disk ($\sim$12 au) by 2~Myr, after which migration slows because only planetesimals previously scattered outward remain available for encounters.

In models 1-4 and 1-6, we set $\mathcal{F}=100$ and $250$, which correspond to $r'_{\mathrm{p}}\simeq700/100=7~\mathrm{km}$ in model 1-4 and $700/250=2.8~\mathrm{km}$ in model 1-6. Compared with models 1-1 to 1-3, outward migration occurs more frequently and often involves multiple embryos that migrate outward in a compact configuration with nearly constant mutual separations. In model 1-4, three embryos migrate monotonically outward to about 9.6, 11.4, and 12.5~au by 1.5~Myr. The two inner embryos then turn inward, whereas the outermost resumes outward migration near 1.9~Myr and reaches around 13.4~au by 2.5~Myr, as shown in the left panel of figure \ref{fig:semi-major1-4_6} and panels a--d of figure \ref{fig:total1-2}. In model 1-6, five embryos exhibit outward migration. Two embryos reach the outer disk edge by 1.9 and 2.5~Myr. Others reverse direction after interactions with neighboring embryos. A low-eccentricity ring with only small embryos embedded forms between about 3 and 5.5~au and acts as a boundary that divides the planetary distribution into inner and outer populations, as seen in the right panel of figure \ref{fig:semi-major1-4_6} and panel h of figure \ref{fig:total1-2}.

Across all six models, including our fiducial model (model 1-5), embryos start growing mostly at radii between 2 and 6 au and then undergo radial dispersion driven by a combination of outward and inward PDM, Type-I migration, and mutual orbital repulsion. Embryos migrating outward frequently migrate in compact packs of a few embryos that maintain nearly constant separation. Within a pack, the outer embryo tends to be the least massive, and the inner ones progressively more massive. In model 1-3, 1-5 and 1-6, the combination of outward and inward PDM, and orbital repulsion produces a low-eccentricity ring with only small embryos embedded and bifurcated distributions of planetary embryos (figure \ref{fig:total}, panel l in figure \ref{fig:total1} and panel h in figure \ref{fig:total1-2}).

\subsection{Results of Model 2: The effect of planetesimal disk size} \label{results:disk_size}
The results of model 1, presented in sections \ref{results:fiducial_model} and \ref{results:gas_drag}, show that protoplanets formed within the inner planetesimal disk migrate outward via PDM to the outer edge of the disk. In what follows, we present the results of model 2, in which the disk is extended to 20 au to investigate the effect of the initial planetesimal disk size. In model 2-1, we increase the number of planetesimals while keeping the initial planetesimal mass, and thus the mass resolution is the same as in model 1. In model 2-2, we further increase the number of planetesimals and reduce the individual mass, thereby improving the mass resolution. We use these setups to investigate the effect of the initial disk size on planet formation, and to assess the robustness of our simulations at higher mass resolution with model 2-2.

In model 2, the planetesimal disk extends from $r_{\mathrm{d,in}} = 2$ au to $r_{\mathrm{d,out}} = 20$ au and is represented by 354,350 planetesimals (model 2-1) and 708,700 planetesimals (model 2-2). For model 2-1, two runs with identical parameters but different random number seeds (Run 1 and Run 2) were performed to examine the dependence on initial conditions. A summary of each model is given in Table~\ref{tab1}.

\begin{figure*}[hbtp]
 \begin{center}
 \includegraphics[width= 15.0cm]{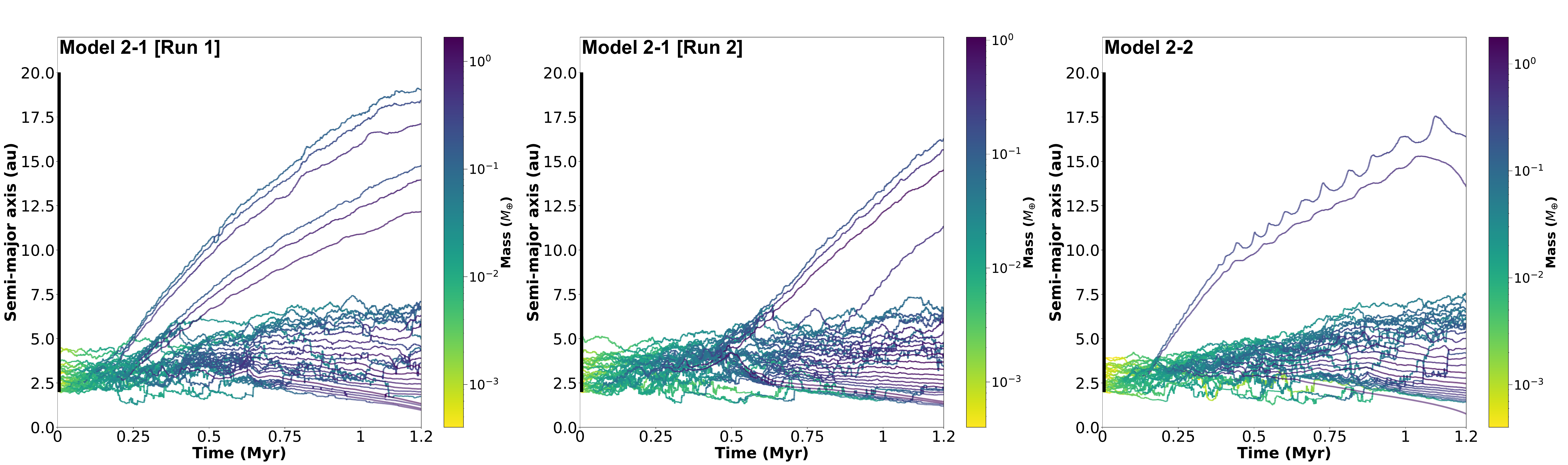}
 \end{center}
 \caption{\textbf{Same as figure \ref{fig:semi-major1-1_3}, but for model run1 of 2-1 (left), run2 of 2-1 (middle), and 2-2 (right).} The vertical axis ranges have been adjusted accordingly (0-22 au). \\{Alt text: The line graphs showing the time evolution of semi-major axes and masses of protoplanets formed in model 2-1 [Run 1], 2-1 [Run 2] and 2-2. The horizontal axis represents time, and the vertical axis displays the semi-major axis in au.}}
 \label{fig:semi-major2-1_2}
\end{figure*}
%

\begin{figure*}[hbtp]
 \begin{center}
 \includegraphics[width= 13.0cm]{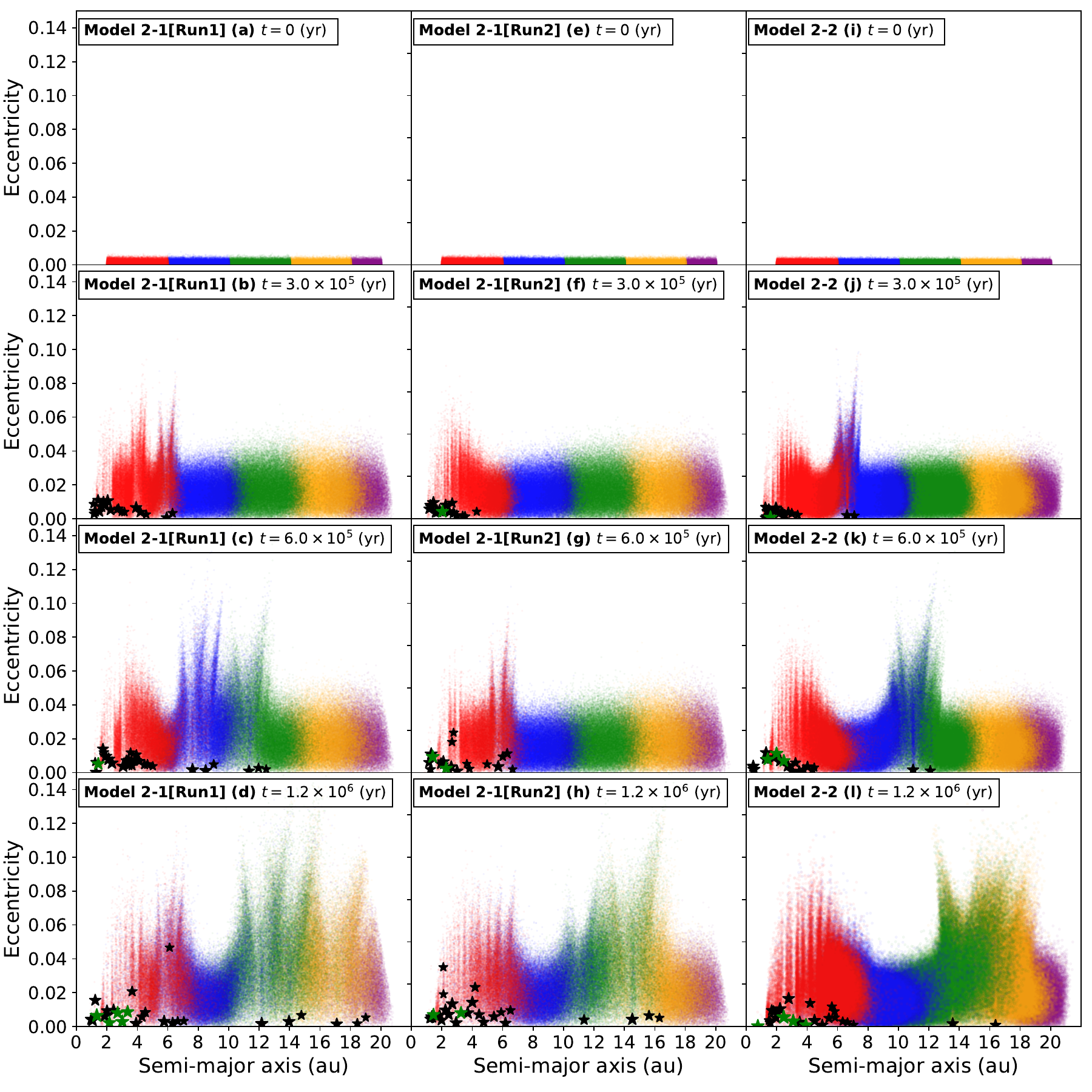}
 \end{center}
 \caption{\textbf{Same as figure \ref{fig:total1}, but showing model run1 of 2-1 (left), run2 of 2-1 (middle), and 2-3 (right).} The horizontal axis ranges have been adjusted accordingly (0-22 au). The dots, colored red, blue, green, yellow, and purple, represent planetesimals. Each color indicates the planetesimals' initial semi-major axes: red for 2-6 au, blue for 6-10 au, green for 10-14 au, yellow for 14-18 au, and purple for 18-20 au. \\ {Alt text: Snapshots of model 2-1 [Run 1], 2-1 [Run 2], and 2-2, arranged chronologically from panel (a) to (d), (e) to (h) and (i) to (l), respectively. The horizontal axis represents semi-major axis in au, and the vertical axis displays eccentricity.}}
 \label{fig:total2}
\end{figure*}

\begin{figure} [hbtp]
  \begin{center}
  \includegraphics[width=9.0cm]{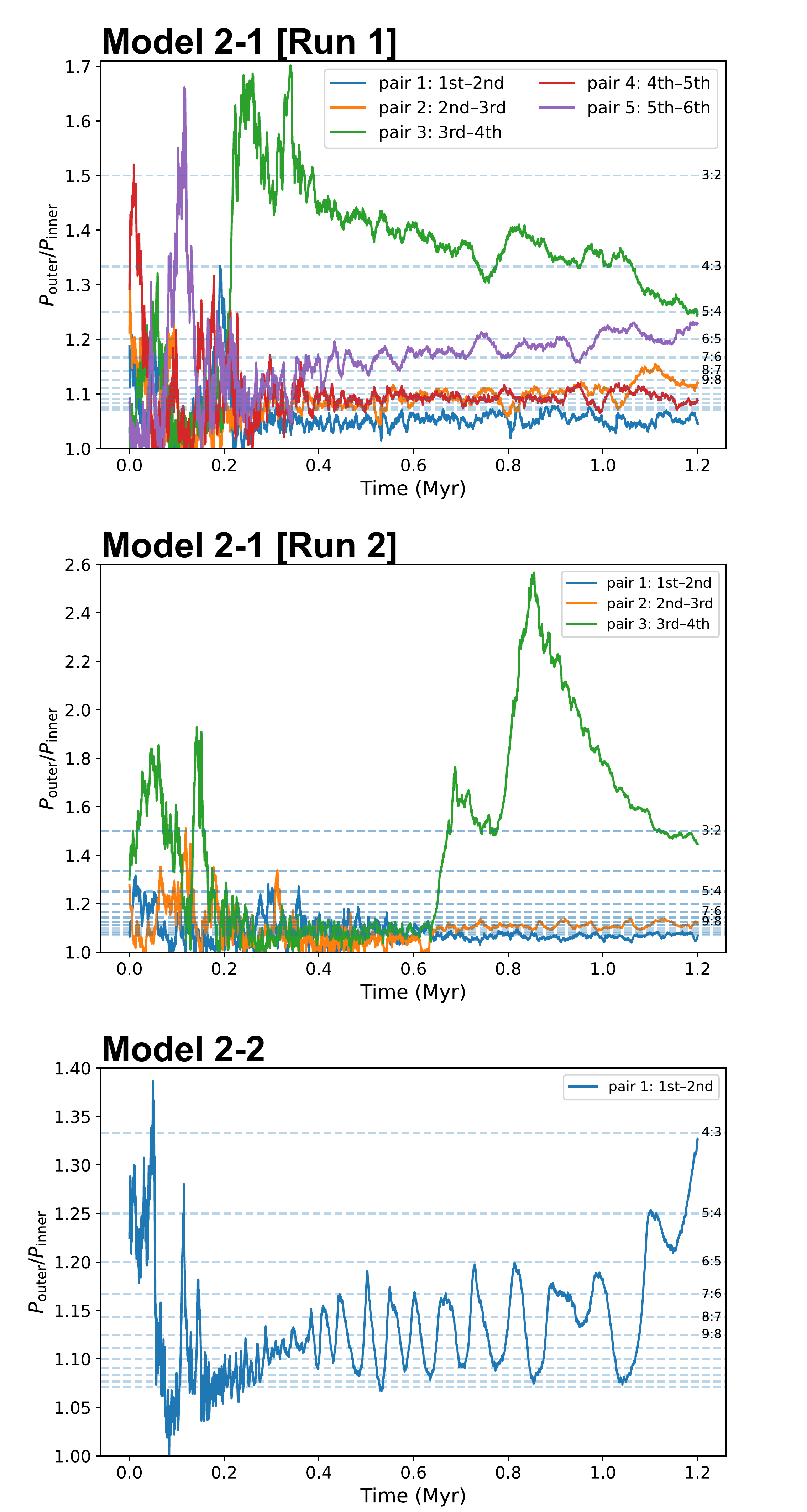}
  \end{center}
  \caption{
  \textbf{The time evolution of the orbital period ratios between neighboring embryos,
  $P_{\mathrm{outer}}/P_{\mathrm{inner}}$, for models 2-1 (runs 1 and 2)
  and 2-2.} In the legend, pair 1 denotes the outermost neighboring pair at the end of the simulation (largest final semi-major axes), pair 2 the next inner pair,
  and so on towards smaller orbital radii. Horizontal dashed lines indicate
  selected first-order resonances $(p+1):p$.\\
  {Alt text: Line graphs showing the time evolution of the mean-motion resonance MMR of outward-migrating embryos in models 2-1 (runs 1 and 2) and 2-2.}}
  \label{fig:MMR}
\end{figure}

Figures \ref{fig:semi-major2-1_2} and \ref{fig:total2} show the orbital evolution of the thirty heaviest planetary embryos formed in model 2 and the snapshots of the entire systems. The overall orbital behavior remains broadly similar to that in model 1: several embryos migrate outward through PDM and eventually reach the outer edge of the disk, around 20 au. The migrating embryos tend to form compact groups that maintain nearly constant orbital separations, consistent with orbital repulsion acting among neighboring embryos. Within each group, the outermost embryo is the smallest, while inner embryos are progressively more massive. When the planetesimal disk is extended to 20 au, several planetary embryos undergo monotonic outward migration through PDM and reach the vicinity of the outer disk edge. This result indicates that outward PDM can operate efficiently even in the outermost regions of a broad planetesimal disk, transporting embryos over a much larger radial distance than in model 1. Another notable feature is that in model 2-2, two embryos exhibit a repetitive, stepwise (``caterpillar-like'') migration pattern, alternately repelling each other while gradually migrating outward (right panel of figure \ref{fig:semi-major2-1_2}). This behavior suggests that mutual gravitational interactions among embryos can modulate the monotonicity of PDM.

To assess whether mean-motion resonances contribute to the compact configurations of outward-migrating embryos, we analyze the time evolution of the orbital period ratios between neighboring embryos in the runs that show clear outward migration. We focus on model 2 for this analysis because, as shown in figures \ref{fig:semi-major1-1_3} and \ref{fig:semi-major1-4_6}, model 1 exhibits frequent orbit crossings and strong scattering among many embryos, so that the ordering of semi-major axes changes repeatedly and neighboring pairs cannot be tracked consistently. In contrast, in model 2, most embryos that undergo outward migration maintain non-crossing, well-ordered orbits up to the end of the simulations (figure \ref{fig:semi-major2-1_2}), which allows us to follow each neighboring pair and calculate the ratio
$P_{\mathrm{outer}}/P_{\mathrm{inner}}$, where $P_{\mathrm{outer}}$ and $P_{\mathrm{inner}}$ denote the orbital periods of the outer and inner embryo in each neighboring pair, respectively, as a function of time (figure \ref{fig:MMR}). In figure
\ref{fig:MMR}, ``pair 1'' denotes the outermost neighboring pair at the end of each simulation (largest final semi-major axes), ``pair 2'' the next inner pair, and so on.

Figure \ref{fig:MMR} shows that neighboring embryos that belong to the same outward-migrating group typically have period ratios in the range $P_{\mathrm{outer}}/P_{\mathrm{inner}} \simeq 1.05$-$1.25$. A notable exception is ``pair 3'' (green curves) in both runs of model 2-1, which connects two distinct groups rather than embryos that migrate outward as a single, coherent group. In model 2-2, only one neighboring pair of embryos migrates outward. For most of the evolution, its period ratio lies in the same range, but it exhibits large oscillations associated with the ``caterpillar-like'' stepwise evolution seen in figure \ref{fig:semi-major2-1_2}. In model 2-1 (run 1), pairs 1, 2, 4, and 5 spend limited intervals near first-order mean-motion resonance ratios. In model 2-1 (run 2), this is most evident for the two outermost neighboring pairs
after $t \simeq 0.65$ Myr, where their period ratios stay close to the 9:8 and 15:14 resonances. These intervals of near-resonant period ratios occur while the embryos are already undergoing outward migration with nearly constant mutual separations. Taken together, these results indicate that the outward migration of the embryos is driven by PDM, while their characteristic spacing is maintained through a combination of mutual
orbital repulsion and occasional, short-lived proximity to first-order mean-motion resonances.

The fact that protoplanets reach the outer edge of the disk, which is set at 12 au in model 1 and extended to 20 au in model 2, implies that if the disk outer edge is located at a greater distance, protoplanets could migrate even farther. Thus, our results suggest that the outer edge of the planetesimal disk in the early solar system must have extended to at least several tens of au to allow for the formation of ice giants. At the same time, our simulations predict that the migration of protoplanets by PDM naturally produces two spatially separated groups of protoplanets: an inner group that originates in the 2-4 au region, and an outer group that migrates to the disk's outer region, separated by an embryo-depleted gap. Future transit surveys such as PLATO \citep{2025ExA....59...26R}, which will provide precise radii, masses, and ages for large samples of small planets around bright Sun-like stars, will be able to test whether analogous bimodal architectures occur in real planetary systems by searching for similar gaps in the radial distribution of low-mass planets.

\section{Discussion and Summary}\label{discussion_summary}

\subsection{Comparison with previous $N$-body simulations of planet formation} \label{discussion:comparison}
In early studies, high-resolution simulations starting from self-gravitating planetesimal disks have been performed (e.g., \cite{2002ApJ...581..666K}). However, owing to computational constraints, these simulations were limited to narrow annular regions near 1 au. Thus, the origin of outer planets and the effect of PDM on planet formation were beyond the scope of these studies. More recent simulations of planet formation with radial extents of more than several au have often assumed the existence of outer planets, or have employed low-resolution, non-self-gravitating planetesimal disks owing to computational limitations (e.g., \cite{2005Natur.435..459T,2005Natur.435..466G,2006Icar..184...39O,2006Icar..183..265R,2011Natur.475..206W,2012A&A...541A..11R,2021Icar..35914305W}). The origin of outer planets and the effect of PDM on planet formation were therefore also beyond the scope of these studies.

In this study, we carried out the first high-resolution, self-consistent $N$-body simulations of planet formation starting from a large-scale, self-gravitating planetesimal disk composed of 237,520 (354,350 or 708,700) planetesimals, initially extending from 2 to 12 au (2 to 20 au). We found that planetary embryos undergo dynamic migration even in the early runaway growth stage from the inner to the outer edges within the conventional protoplanetary disk. Such active migration serves as a rapid radial diffusion mechanism for protoplanets and significantly influences the early stages of planetary formation.

\subsection{Implications of observed disk structures and modern disk models for planet formation} \label{discussion}
Here, we discuss the implications of recent observations of protoplanetary disks and modern disk evolution models for the planet formation process, in light of our results. In this study, we adopted an MMSN-like disk model that has been widely used in many studies (e.g., \cite{2002ApJ...581..666K,2014Icar..232..118M,2024PASJ...76.1309J}). We limited our study to the MMSN disk model due to the limitation of computational resources associated with the high computational cost of our large-scale simulation, which fully accounts for the gravitational interactions among more than $2 \times 10^5$ planetesimals. However, recent dust observations in protoplanetary disks suggest that the radial profile of dust surface density follows $\Sigma\propto r^{-1}$ or an even shallower slope \citep{2009ApJ...700.1502A}. Compared to the MMSN model, such disks have a more extended dust distribution toward the outer regions. As a result, the region in which PDM-driven outward migration is most effective is expected to extend beyond the range covered in our simulations. Furthermore, in \authorcite{2009Icar..199..197K} (\yearcite{2009Icar..199..197K}), it was shown that the frequency of outward migration due to PDM increases as the slope of the surface density profile becomes shallower. This implies that solid material can be transported more efficiently to the outer disk in disk models with shallow surface density profiles. Compared to the MMSN-like disk model used in this study, such disks are expected to facilitate even more dynamic planetary migration.

Moreover, recent observations of protoplanetary disks have also revealed that disks often exhibit rings and gaps in the dust continuum, indicating that the distribution of solids is not always smooth but often structured in radius (e.g., \cite{2015ApJ...808L...3A,2018ApJ...869L..41A}). From a theoretical point of view, a broad class of models suggests that such rings arise at locations where the disk structure changes and the solid surface density is locally enhanced, for example near the inner edge of dead zones, snow lines, or sublimation lines (e.g., \cite{2014ApJ...780...53C,2022NatAs...6..357I,2022A&A...660A.117H}). At these locations, variations in gas pressure, ionization state, or particle size can lead to efficient trapping of dust and pebbles and to the formation of narrow rings with locally enhanced solid surface density. The subsequent formation and growth of planetesimals and planetary embryos in such rings, and their dynamical evolution, have been investigated in a number of studies (e.g., \cite{2023NatAs...7..330B,2023PASJ...75..951J,2024ApJ...972..181O,2025ApJ...985...71K}). Compared to the smooth MMSN-like disk considered in this study, such ringed planetesimal disks have a much more radially concentrated dust distribution. In such disks, the discrete rings may act as quasi-isolated source regions for planetary building blocks. In this case, PDM would still operate and could efficiently stir and spread embryos within each ring and exchange material between neighboring rings, but the large-scale redistribution of solids would be partly counteracted by trapping at pressure maxima associated with the rings. The net redistribution is expected to be governed by the competition between PDM-driven spreading and trapping at such substructures.

In addition, modern disk evolution models based on MHD simulations have shown that magnetically driven disk winds combined with viscous diffusion can modify the gas surface density profile of protoplanetary
disks and even produce regions in the inner disk where the surface
density increases with radius (i.e., a positive radial gradient) 
\citep{2016A&A...596A..74S}. In such disks, the gas density in the inner region can be substantially reduced compared to MMSN, which affects both the efficiency of Type-I migration and the coupling between solids and gas \citep{2017A&A...608A..74O,2018A&A...615A..63O,2018A&A...612L...5O}. Our present MMSN-like model, therefore, provides a useful baseline for understanding how PDM operates in a classical smooth disk, but it is important to examine how our conclusions change in more realistic, wind-driven disk models. We plan to extend our large-scale, self-consistent simulations to disks with surface density profiles consistent with such modern disk evolution models and with observations of protoplanetary disks, in order to further investigate the impact of planetary migration on the planet formation process as a next step.

\subsection{The effect of fragmentation on PDM} \label{discussion:fragment}
In this study, we adopted a perfect accretion model in which all collisions between particles result in complete merging. Including fragmentation could lead to quantitative changes in our results. In our simulations, planetesimals exterior to an outward-migrating embryo can grow efficiently by perfect accretion. If fragmentation were taken into account, however, small fragments would be continuously replenished outside the embryo. These small fragments are more strongly affected by gas drag and therefore tend to have lower random velocities. As a consequence, fragmentation would likely facilitate PDM by maintaining a reservoir of low-velocity scatterers exterior to the embryo. A self-consistent treatment of fragmentation is computationally expensive and is therefore left for future work.

\subsection{Summary} \label{summary}
In this study, we performed the first high-resolution, self-consistent $N$-body simulations of planet formation starting from a large-scale, self-gravitating planetesimal disk. These simulations were carried out on the supercomputer Fugaku. Our main findings are summarized as follows.
\begin{enumerate}
    \item PDM acts as a rapid radial diffusion mechanism for planetary embryos from the early runaway growth stage, driving both inward and outward migration of bodies initially formed in the inner disk (figures \ref{fig:semimajor_grid}-\ref{fig:total2}).
    \item Migrating embryos tend to form compact groups with nearly constant separation, consistent with orbital repulsion. Within such groups, the outermost body is typically the smallest, and inner bodies are progressively more massive (figures \ref{fig:semimajor_grid}, \ref{fig:semi-major1-1_3}, \ref{fig:semi-major1-4_6}, and \ref{fig:semi-major2-1_2}). Taken together with the resonance analysis in section \ref{results:disk_size} and figure \ref{fig:MMR}, these results suggest that the outward migration of embryos is driven by PDM, while their characteristic spacing is maintained through a combination of mutual orbital repulsion and mean-motion resonances.
    \item Some embryos exhibit a repetitive, stepwise (``caterpillar-like'') migration, indicating that mutual gravitational interactions can modulate the monotonicity of PDM (figure \ref{fig:semi-major2-1_2}).
    \item A bimodal embryo configuration with an embryo-depleted gap has been observed in more than half of the runs. This gap reflects the combined effects of outward PDM, inward PDM, and orbital repulsion (figures \ref{fig:semimajor_grid}-\ref{fig:total2}).
    \item Regardless of whether the disk spans 2-12 au (model 1) or 2-20 au (model 2), embryos migrate to the outer edge. This indicates that, in a more extended planetesimal disk, PDM can deliver solids to even larger radii and facilitate the growth of the ice giants' cores.
\end{enumerate}
Taken together, these results support a dynamic formation pathway in which planetary embryos grow while migrating over large radial distances (figure \ref{fig:schematic}b), rather than forming strictly in-situ at their initial locations (figure \ref{fig:schematic}a). Our findings challenge the traditional view of ``in-situ'' planet formation scenarios and mark a significant step forward in our understanding of planetary formation processes.



\begin{ack}
We are very grateful to an anonymous referee for providing useful feedback that contributed to the improvement of our paper. We would like to thank Eiichiro Kokubo and Yuji Matsumoto for their helpful advice throughout this work. This work was supported by MEXT as the ``Program for Promoting Researches on the Supercomputer Fugaku'' (Structure and Evolution of the Universe Unraveled by Fusion of Simulation and AI; Grant Number JPMXP1020230406) and JSPS KAKENHI Grant Number 25KJ1814. Computational resources were provided by the RIKEN Center for Computational Science through the use of the supercomputer Fugaku (Project IDs: hp240219, hp240094 and hp250056). Test simulations in this paper were also carried out on a Cray XC50 and XD2000 systems at the Centre for Computational Astrophysics (CfCA) of the National Astronomical Observatory of Japan (NAOJ).
\end{ack}

\section*{Funding}
This work was supported by MEXT as the ``Program for Promoting Researches on the Supercomputer Fugaku'' (Structure and Evolution of the Universe Unraveled by Fusion of Simulation and AI; Grant Number JPMXP1020230406) and JSPS KAKENHI Grant Number 25KJ1814.

\section*{Data availability} 
The data that support the findings of this study are available from the corresponding author on request.

\appendix 

\section{The validation of mass resolution correction using the gas drag coefficient} \label{appendix}
\begin{figure*}[hbtp]
 \begin{center}
 \includegraphics[width= 14.0cm]{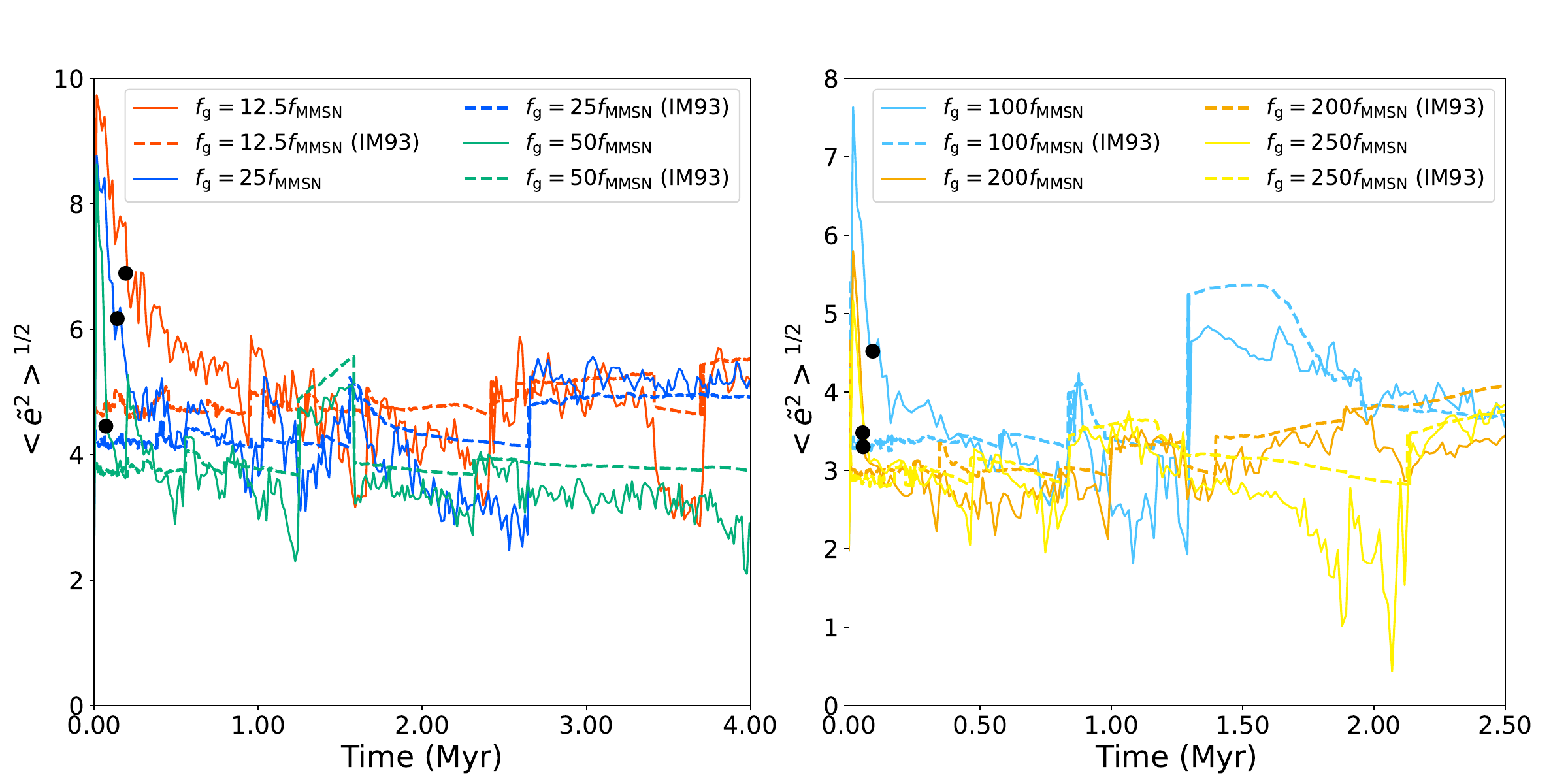}
 \end{center}
 \caption{\textbf{The time evolution of the RMS eccentricity of planetesimals within the feeding zones ($a_{\mathrm{p}} \pm 5r_{\mathrm{Hill}}$) of the largest protoplanets in the planetesimal disk for models 1-1 to 1-6 (solid curves).} The dashed curves represent the estimated RMS eccentricity values calculated for $r_{\mathrm{p}}'=[1/12.5, 1/25, 1/50, 1/100, 1/200, 1/250]r_{\mathrm{p}}$ using equation (\ref{eq:eccentricity1}). Black dots on the solid curves indicate when the mass ratio between the protoplanet and the planetesimals exceeds 100.\\{Alt text: Line graphs showing RMS eccentricity evolution of planetesimals around the largest protoplanets in models 1-1 to 1-6 (solid curves), with analytic estimates (dashed curves) and black dots mark where the protoplanet-planetesimal mass ratio exceeds 100.}}
 \label{fig:proto_eccentricty}
\end{figure*}

In this section, we compare our simulation results with the early study of \cite{1993Icar..106..210I} (hereafter IM93), which examined the dynamical evolution of planetesimals around protoplanets using semi-analytical and $N$-body simulations. Based on this comparison, we discuss the validity of using the gas drag coefficient as a free parameter to represent planetesimal disks with various planetesimal sizes.

According to IM93, beyond the snowline, the effects of dynamical friction dominate the dynamical evolution of planetesimals around protoplanets from as early as when the mass ratio between the protoplanet and planetesimals slightly exceeds 100. The averaged eccentricity and inclination of planetesimals near a protoplanet at this stage are estimated by IM93 as
\begin{equation}
\langle\tilde{e}^2\rangle^{1/2}=2\langle\tilde{i}^2\rangle^{1/2}=6(\tilde{\tau}_{\mathrm{gas}}/1000)^{1/6}. \label{eq:eccentricity1}
\end{equation}
Here $\tilde{e}$, $\tilde{i}$, and $\tilde{\tau}_{\mathrm{gas}}$ represent the normalized eccentricity, inclination, and the characteristic timescale of gas drag, respectively, given by:
\begin{eqnarray}
\tilde{e}&=&e/h,\\
\tilde{i}&=&i/h,\\
\tilde{\tau}_{\mathrm{gas}}&=&\tau_{\mathrm{gas}}/{T_{\mathrm{K}}}=\frac{2m}{C_{\mathrm{D}}\pi r_{\mathrm{p}}^2\rho_{\mathrm{gas}}v_{\mathrm{K}}T_{\mathrm{K}}}\label{eq:tau}, 
\end{eqnarray}
From equations (\ref{eq:eccentricity1}) to (\ref{eq:tau}), the averaged eccentricity $\langle\tilde{e}^2\rangle^{1/2}$ depends weakly on the disk gas density $\rho_{\mathrm{gas}}$ and the planetesimal radius $r_{\mathrm{p}}$. This is because the characteristic timescale of gas drag $\tilde{\tau}_{\mathrm{gas}}$ is inversely proportional to the disk gas density and directly proportional to the planetesimal radius, assuming constant internal density and semi-major axis. Thus, altering the gas drag coefficient by a factor of $k$ is equivalent to changing the radius of planetesimals within the feeding zone of protoplanets by a factor of $k^{-1}$ when considering the dynamical evolution of the planetesimals.

Figure \ref{fig:proto_eccentricty} shows the RMS eccentricity of planetesimals within the feeding zones ($a_{\mathrm{p}} \pm 5r_{\mathrm{Hill}}$) of the largest protoplanets in the planetesimal disk (2 au $< r <$ 12 au) for models 1-1 to 1-6 (solid curves) and the estimated RMS eccentricity values, calculated for $r_{\mathrm{p}}'=[1/12.5, 1/25, 1/50, 1/100, 1/200, 1/250]r_{\mathrm{p}}$ by using equation (\ref{eq:eccentricity1}) (dashed curves). Each black dot on the solid curves represents the time when the mass ratio between the protoplanet and the planetesimals exceeds 100. Figure \ref{fig:proto_eccentricty} indicates that as the gravitational scattering by the protoplanet becomes dominant, the RMS eccentricity of planetesimals within the feeding zone aligns well with the estimates given by IM93 in all models. Thus, we can assume that models 1-1 to 1-6, which apply an enhancement parameter for gas density of $\mathcal{F}=[12.5, 25, 50, 100, 200, 250]$, effectively represent the dynamical evolution of planetesimals with sizes of $r_{\mathrm{p}}'=[1/12.5, 1/25, 1/50, 1/100, 1/200, 1/250]r_{\mathrm{p}}$ under gas drag conditions similar to those in the MMSN.

Here, we should note that this argument is only a rough estimate intended to demonstrate the validity of varying the gas drag coefficient as a parameter. To precisely discuss whether changing the gas drag coefficient can accurately reproduce disks with various planetesimal sizes, large-scale $N$-body simulations that fully reproduce realistic planetesimal disks are necessary\footnote{Assuming the MMSN, we need a huge number of particles of at least $N \sim 10^7$ to self-consistently reproduce realistic planetesimal sizes (10 km to $10^2$ km) within a few au-wide planetesimal disk in the Jovian region. Currently, we are conducting self-consistent, large-scale $N$-body simulations of planet formation using $N = 10^4$ to $N = 10^8$ on the supercomputer Fugaku to thoroughly investigate the dependence on mass resolution (Jinno et al., in preparation). By comparing the results of this ongoing research with those presented in this paper, we can precisely validate whether varying the gas drag coefficient as a parameter can accurately replicate disks with diverse planetesimal sizes.}.

\section{Comparison of migration timescales of PDM and Type-I migration} \label{appendix2}
Here we compare the characteristic migration timescales associated with PDM and Type-I migration for the representative protoplanets I-III for our fiducial model, model 1-5.

Following the fiducial analytic expression for the PDM migration rate \citep{2000ApJ...534..428I,2009Icar..199..197K}, we define an effective PDM migration timescale as
\begin{equation}
 \tau_{\mathrm{PDM}}
 \equiv
 \frac{a_{\mathrm{p}}}{\left|\mathrm{d}a_{\mathrm{p}}/\mathrm{d}t\right|_{\mathrm{fid}}}
 \approx
 \frac{M_* T_{\mathrm{p}}}{4\pi \Sigma_{\mathrm{dust}} a_{\mathrm{p}}^2},
 \label{eq:tPDM_def}
\end{equation}
where $a_{\mathrm{p}}$ is the semimajor axis of the protoplanet, $M_*$ is the
stellar mass, and $T_{\mathrm{p}}$ is the orbital period of the protoplanet. The quantity
$\Sigma_{\mathrm{dust}}$ denotes the local surface density of planetesimals
around the protoplanet, which we estimate directly from the simulations by
summing the masses of planetesimals located within $\pm 5 r_{\mathrm{Hill}}$ of the
protoplanet and dividing by the area of the corresponding annulus.

For Type-I migration, we use equation (\ref{eq:tau_a}), and we combine it with the PDM migration timescale $\tau_{\mathrm{PDM}}$ to evaluate the ratio $\tau_{\mathrm{TypeI}}/\tau_{\mathrm{PDM}}$ for protoplanets I-III. By combining
equation (12) with equation~(\ref{eq:tPDM_def}), we obtain
\begin{equation}
 \frac{\tau_{\mathrm{TypeI}}}{\tau_{\mathrm{PDM}}}
 =
 \frac{4\pi}{C_{\mathrm{T}}}\,
 h^{2}\,
 \frac{\Sigma_{\mathrm{dust}}}{\Sigma_{\mathrm{gas}}}\,
 \frac{M_*}{M}\,
 \frac{1}{T_{\mathrm{p}}\,\Omega_{\mathrm{K}}}\,
 \left[
  1 + \frac{C_{\mathrm{T}}}{C_{\mathrm{M}}}
      \left(\hat{e}^2 + \hat{i}^2\right)^{1/2}
 \right],
 \label{eq:tau_over_tPDM}
\end{equation}
the coefficients $C_{\mathrm{T}}$ and $C_{\mathrm{M}}$ are the same as those
used in equation (\ref{eq:tau_a}).

\begin{figure}[hbtp]
 \begin{center}
 \includegraphics[width= 8.0cm]{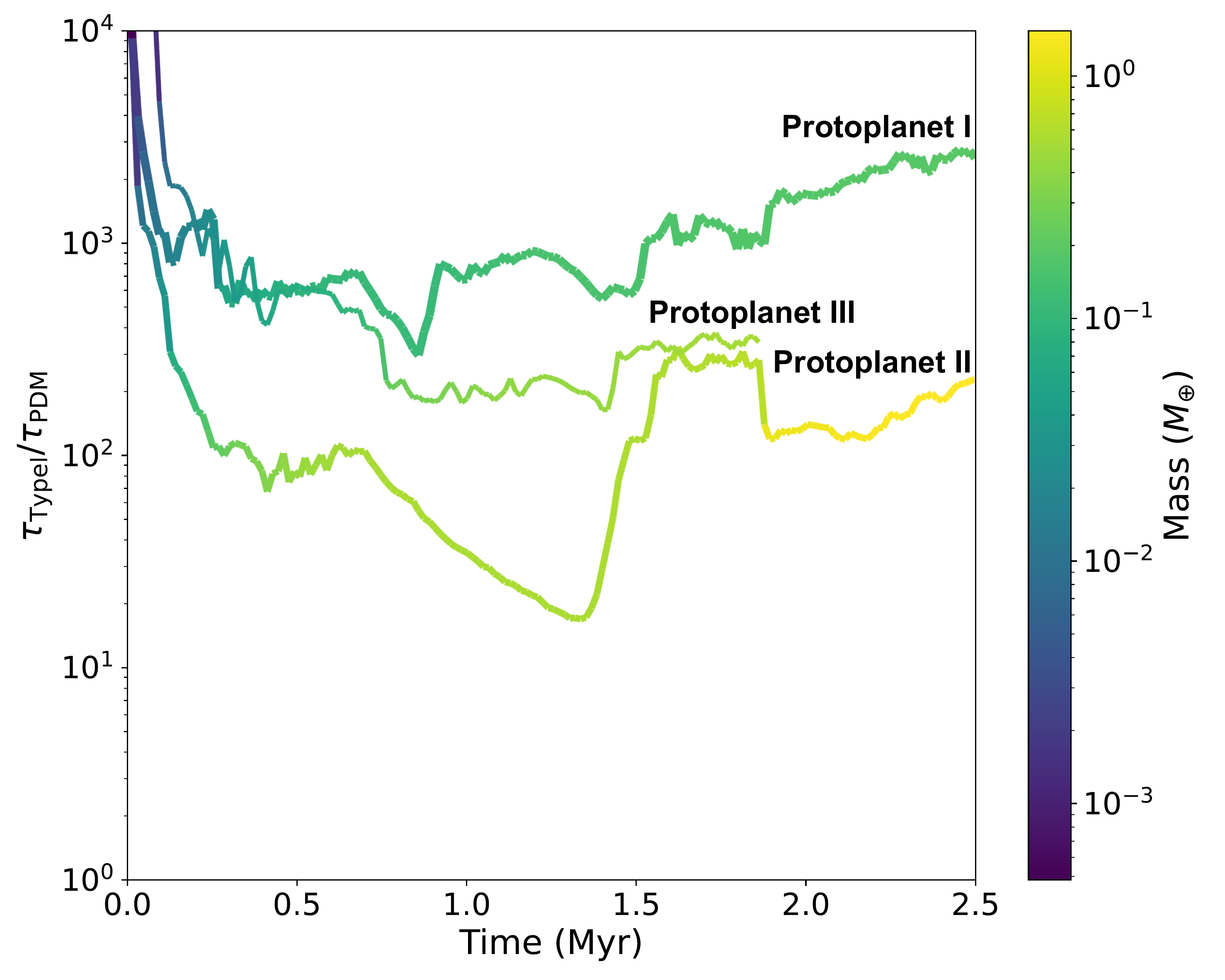}
 \end{center}
 \caption{\textbf{Time evolution of the ratio $\tau_{\mathrm{TypeI}}/t_{\mathrm{PDM}}$ for protoplanets~I-III. The color along each curve indicates the planetary mass}\\{Alt text: The line graph showing the ratio $\tau_{\mathrm{TypeI}}/\tau_{\mathrm{PDM}}$ evolution of protoplanets~I-III in models 1-5.}}
 \label{fig:appendix_timescale}
\end{figure}

We evaluate equation~(\ref{eq:tau_over_tPDM}) for protoplanets I-III as a
function of time, using the local values of $\Sigma_{\mathrm{dust}}$, $\Sigma_{\mathrm{gas}}$, $M_{\mathrm{p}}$, $a_{\mathrm{p}}$,
$e$, $i$, $h$, and $\Omega_{\mathrm{K}}$ taken from our simulations. The resulting time evolution of $\tau_{\mathrm{TypeI}}/\tau_{\mathrm{PDM}}$ is shown in figure~\ref{fig:appendix_timescale}. During the phases in which the protoplanets undergo migration reversal (at 0.85 Myr for protoplanet~I and 1.35 Myr for protoplanet~II), the ratio $\tau_{\mathrm{TypeI}}/\tau_{\mathrm{PDM}}$ typically lies in the range $\sim 10$-$100$, indicating that the PDM timescale is shorter than the Type-I migration timescale by about one to two orders of magnitude.

We should note that both $\tau_{\mathrm{TypeI}}$ and $\tau_{\mathrm{PDM}}$ should be
regarded as order-of-magnitude estimates. The fiducial PDM rate
(\ref{eq:tPDM_def}) is derived for an isolated protoplanet migrating in a
locally smooth planetesimal disk, whereas in our simulations the local
planetesimal distribution is modified by depletion of planetesimals and the
presence of multiple neighboring embryos. These effects can lengthen the
effective PDM migration timescale by a factor of a few or more. Similarly, the Type-I
prescription in equation (\ref{eq:tau_a}) assumes an isolated planet embedded in a smooth gas disk and does not include the perturbations from neighboring planetesimals or embryos. Therefore, the absolute values of $\tau_{\mathrm{TypeI}}$ and $\tau_{\mathrm{PDM}}$ are uncertain at the level of at least a factor of a few, but the relative ordering and the fact that PDM operates more rapidly than Type-I migration are robust.

\bibliographystyle{aasjournal}
\bibliography{reference}
\end{document}